\documentclass[numbered]{myclass}
\usepackage{graphicx}
\usepackage{ccaption}
\usepackage{subcaption}
\usepackage[hidelinks]{hyperref}
\usepackage{cleveref}

\AuthorHeaders{Guo, Martínez, Correia, and Van Arem}
\title{Impact of Trip Distance Distribution Time Dependency and Aggregation Levels in Bathtub Models - A Comparative Simulation Analysis}

\author{%
  \textbf{Jiayi Guo}\\
  Delft University of Technology\\
  Department of Transport and Planning\\
  Stevinweg 1, 2628 CN, Delft, the Netherlands\\
  Tel: +31 682530634; Email: J.Guo-5@tudelft.nl \\
  Corresponding author\\
  \hfill\break% this is a way to add line numbering on empty line
  \textbf{Irene Martínez}\\
  Delft University of Technology\\
  Department of Transport and Planning\\
  Stevinweg 1, 2628 CN, Delft, the Netherlands\\
  Email: I.Martinez@tudelft.nl\\
  \hfill\break%
  \textbf{Gonçalo Correia}\\
  Delft University of Technology\\
  Department of Transport and Planning\\
  Stevinweg 1, 2628 CN, Delft, the Netherlands\\
  Email: G.Correia@tudelft.nl\\
  \hfill\break%
  \textbf{Bart van Arem}\\
  Delft University of Technology\\
  Department of Transport and Planning\\
  Stevinweg 1, 2628 CN, Delft, the Netherlands\\
  Email: B.vanArem@tudelft.nl
}

\begin{document}
\maketitle

\section{Abstract}
Bathtub models are used to study urban traffic within a certain area. They do not require to take into account the detailed network topology. The emergence of different bathtub models has raised the question of which model can provide more robust and accurate results under different demand scenarios and network properties. This paper presents a comparative simulation analysis of the accumulation-based model and trip-based models under static and dynamic trip distance distribution (TDD) scenarios. Network accumulation was used to validate and compare the performance of the bathtub models with results from the macroscopic traffic simulation with dynamic traffic assignment. Three networks were built to explore the effect of network properties on the accuracy of bathtub models. Two are from the network of Delft, the Netherlands, and one is a reference toy network. The findings show that the time dependency of TDD can increase the errors in bathtub models. Using TDD in different aggregation levels can significantly influence the performance of bathtub models during demand transition periods. The state transition speed of networks is also found to be influential. Future research could explore the effects of dynamic TDD under congested situations and develop enhanced bathtub models that can better account for different network state transition speeds.  \\ \\ 
206 words

% Lastly, network properties such as road hierarchy, the spatial distribution of demand, and the existence of bottlenecks are found to be influential on the performance of MFD-based models. These factors should be considered in case studies and control strategy design for cities.  \\ \\
% The Transportation Research Board (TRB) has unique and seemingly arbitrary requirements for manuscripts submitted for review. These requirements make it difficult to write the manuscripts quickly, and no existing \LaTeX\ style comes close to fooling the guidelines. This represents an initial effort at creating a template to meet the requirements of TRB authors using \LaTeX, R, Sweave, and/or other literate programming software.
\hfill\break%
\noindent\textit{Keywords}: Macroscopic Fundamental Diagram, Bathtub model, Trip distance distribution, Trip-based model, dynamic traffic assignment.
\newpage
\section{Introduction}
% \textbf{MFD(Macroscopic Fundamental Diagram)-based models}\label{sec:litreview_models}\\ \\
Urban traffic congestion has been causing a massive amount of time loss worldwide. According to \citep{tomtomindex}, more than 50 cities worldwide are facing more than 1.5 times free-flow travel time during rush hours. \citep{INRIX2017} estimated that congestion hotspots in 123 European cities would result in an economic loss of 217 billion euros. To tackle issues caused by the increasing travel demand, more regional and network-level traffic management strategies are required to improve the efficiency of urban networks. However, topology-based macroscopic and microscopic models are computationally expensive to build and calibrate. Some of the details in these models are not necessary to assess network-level traffic control, such as congestion pricing and perimeter control. 

In this context, a more aggregate and flexible model without considering topological structure has gained attention among researchers over the past two decades \cite{Johari2021}. This model assumes a fundamental relationship between the aggregated vehicle accumulation and the network flow, called the Macroscopic Fundamental Diagram (MFD). The existence of MFD was proven from empirical data in various cities worldwide \cite{Geroliminis2008, lu-2018, Paipuri2020, ambuhl-2021}. With this model, the network dynamics can be modeled at an aggregated level, where the network is assumed to be a reservoir (a.k.a. bathtub) governed by the conservation of traveled distance and the MFD. This group of models is known as MFD-based models or bathtub models \cite{Johari2021}. In this study, we refer to the model simply as the "bathtub model".
% problem, background, why MFD, scope of this kind of questions
Different bathtub models have been proposed in the literature, and there is no common view on which formulation is better for modeling urban traffic. Therefore, it continues to be very relevant and necessary to understand the differences between the different bathtub models and compare them, especially under different network and demand properties. 
% \subsection{MFD-based models}
% from 2007 (the year when \cite{Daganzo2007} presented the accumulation-based model) and on, is the one in which the idea of macroscopic urban traffic network modeling and control is widely researched. The general idea is to ignore the topological structure of the road network and consider it as a bathtub/reservoir with a certain relation between vehicle speed/outflow and density within the area. That is, the speed $V(t)$ is a function of the accumulation ($n(t)$) or density ($k(t)$) in the network.
% \begin{equation}
%     V(t) = V(n(t)) \quad or \quad V(t) = V(k(t))
%     \label{eq:mfd}
% \end{equation}
% These studies often focus on an area of a city, which has a size within 10$km^2$ for single-reservoir research (4$km^2$ in Zurich \cite{Mariotte2017}, 5$km^2$ in San Francisco \cite{Daganzo2008}, and 10$km^2$ in Yokohama \cite{Geroliminis2008}). For multi-reservoir research, the total area ranges from 2.5$km^2$ with 3 reservoirs \cite{Ji2012} to 80$km^2$ with 5 and 10 reservoirs \cite{Mariotte2020case} and 200$km^2$ with 14 reservoirs\cite{Fu2020}. In multi-reservoir studies that focus on a larger area, the sizes of individual reservoirs are also larger than the size of reservoir in single-reservoir studies. \\ \\

The accumulation-based model is the most widely used bathtub model for macroscopic urban network modeling. The origin of this model was in an unpublished paper by Vickrey in 1991, which was found later and published in 2020 \cite{Vickrey2020}. This model is also known as Vickrey's bathtub model. This paper derived an ordinary differential equation (ODE) \cite{Andrew1976} for network outflow expression from network accumulation conservation. This model assumes that the trip distances of the vehicles inside the modeled area (the bathtub) follow a negative exponential (NE) distribution. 
\citep{Small2003} and \citep{Daganzo2007} also used ordinary differential equations to describe the dynamics of the number of vehicles within an area. \citep{Small2003} did not clarify the trip distance assumption that they used. \citep{Daganzo2007} assumed the same trip distance for all origins in a conceptual ring network. Since \cite{Daganzo2007}, accumulation-based models were considered to assume the same trip distance for all vehicles until the work of \citep{Vickrey2020} was published. 

Although the accumulation-based model is convenient to implement, it has two main limitations: (1) The outflow expression is only valid when the trip distances follow an NE trip distance distribution (TDD); (2) The model is memory-less. That is, the outflow depends only on the accumulation of the current time step, regardless of how far trips have traveled in the past. The first limitation may lead to inaccuracies since empirical data from cities show different types of distributions than NE distribution \cite{colak-2016, yang-2018, Martinez2021}. Whilst the second limitation results in inaccurate travel time during demand changes \cite{Mariotte2017, Leclercq2019}. Both limitations restrict the scenarios in which the accumulation-based model can be used as a reliable prediction for aggregated network traffic dynamics. 

To address the second drawback, \citep{Haddad2020} introduced the accumulation-based model with delayed MFD outputs, which includes travel time in single-region and multi-region cases as a system delay. However, the delay of the network is assumed to be the same during the duration of trips, while the network accumulation can vary during that period. \citep{Huang-2020} proposed an extended model where the delay term is calculated based on the network average speed in every time step. Accumulation-based models with delayed MFD outputs can be used only when the First-In-First-Out (FIFO) principle holds (i.e., all vehicles have the same trip distance), thus they still have the limitation of requiring a specific TDD type.

The trip-based model addresses the two drawbacks by considering individual trips instead of the accumulation. \citep{Arnott2013} firstly mentioned an expression of trip distance with the trip-based modeling idea. \citep{Mariotte2017} and \citep{Lamotte2018} introduced the formulation of the trip-based model. The former paper proposed two simulation strategies for the trip-based model. The difference between them is whether FIFO is required. The latter paper used a trip-based simulation to explore the equilibrium characteristics during the morning peak. Although the trip-based model can solve the two drawbacks of the accumulation-based model, it is more computationally expensive. The computational complexity of trip-based models can be reduced by leveraging the efficient sorting of trips and down-scaling methods \cite{martinez-2024}.

Alternatively, some researchers aimed to capture trip distance heterogeneity while considering aggregated or continuous demand. There are two continuum models that aim to better capture how trips with different trip distances progress in the system, namely, the M model and the generalized bathtub model. M model \cite{Lamotte2018-2, Sirmatel2021} considers the conservation of the total remaining distance. \citep{jin-2020} presented the generalized bathtub model, which introduces the conservation of vehicle accumulation with different remaining trip distances. The most essential extension made by M, trip-based, and generalized bathtub models is that the outflow expression is valid for different TDD types.

% The features of different types of MFD-based models are listed in Table \ref{tab:TDD assumption}. Except for the trip-based model, every other method considers continuous demand and uses either ODE or PDE for the state update functions. Accumulation-based and delay-based models rely on the assumption of a negative exponential TDD, while the trip-based model, M model, and GBM can handle various distribution types. From the perspective of TDD information usage, these 5 models can be divided to three different categories. Assuming TDD data are collected in histograms, trip-based model and GBM can utilise all the attribute and details. The former uses the trip distance of each individual trips, and the latter can support any type of cumulative distribution functions. The second category contains the M model. It can use the aggregated mean and variance of the collected TDD. As for the accumulation-based and delay-based model, they can only process the aggregated average trip distance, as the distribution type is assumed. It can be expected that with more complex and atypical trip distance data, the error will be larger when less information is considered in the model. \\ \\

% \subsection{Model comparison studies}
%2017(trip-based), 2019(fast-varying),2019(flow exchange),2021(estimation) m model control, bi-model one, and the one in 2024. 

Few previous studies have focused on comparing different types of bathtub models. \citep{Mariotte2017} presented solution processes for accumulation-based and trip-based models, where the trip-based model can be solved numerically using either fixed time steps or fixed vehicle increments. The latter, known as the event-based method, updates the simulation at each trip entry and exit. The accumulation-based model showed a slow response under fast-varying demand, which is a direct result of the model's “lack of memory of the past” and “mostly ignores the traveled trip distances” according to the \citep{Mariotte2017}. Fast-varying demand was then further discussed in \citep{Leclercq2019}, which compared the performance of accumulation-based and trip-based models with microsimulation, assuming homogeneous trip distances. The accumulation-based model better captures the saturation case (when the traffic demand is close to the capacity of the network), while the trip-based model is more accurate during the free-flow condition. Trip-based simulation with fixed time steps in \cite{Mariotte2017} overestimates the accumulation during the saturation state. Similar results were also observed by \citep{Mariotte2019} and \citep{huang-2024}.
\citep{Sirmatel2021} compared the accumulation-based model and M model with microsimulation. Since the TDD data does not follow NE distribution, outflow expression in the accumulation-based model is less accurate, leading to larger error compared with the M model. \citep{Paipuri2020_bimodal} compared the accumulation-based, trip-based, and delay-based models under a low-demand scenario. The trip-based simulation has the closest results compared to the micro-simulation. 

% \citep{huang-2024} compared the accumulation-based model, trip-based model, and accumulation-based model with delayed MFD outputs in single-region and multi-region cases. Comparing model results with microsimulation results, they observed the inaccuracy of the trip-based model during saturation states, as others have observed previously \cite{Mariotte2019, Leclercq2019}. 

In summary, previous studies have concluded that specific extensions of accumulation-based models, such as state-dependent delays or total remaining distance, can mitigate the drawbacks of having no memory of the past. The trip-based model is more accurate than the accumulation-based model in free-flow states compared to microscopic traffic simulations, but it is less accurate in saturation states. These conclusions were reached under the assumption of homogeneous trip distances. \citep{Leclercq2019}, \citep{Sirmatel2021}, and \citep{huang-2024} used OD data from areas in cities such as Lyon, Barcelona, and Hangzhou in their microsimulation. Only the mean trip distance was used as input for the bathtub models, instead of considering the entire distribution. 

We argue that using only a constant and homogeneous trip distance for the trip-based model in comparison is an unnecessary simplification since an important advantage of the trip-based model is its ability to handle various TDD types and individual trip distances. The study by \citep{paipuri-2019} compared the accumulation-based model and the trip-based model using different aggregations of TDD input. For the trip-based model, using trip distance groups and individual trip distances showed lower errors than using more aggregated TDD input. This supports the claim that utilizing less aggregated TDD information is an advantage of the trip-based model. \citep{paipuri-2019} also found that using the mean of the trip distance (MTD) led to the lowest error for the accumulation-based model. In the following, we refer to the different aggregations of TDD information as the aggregation level of TDD. The errors are calculated using microscopic traffic simulation as the reference. Although the difference in numerical results was compared, it is unclear whether the added error when using MTD for every trip comes from the demand change period or steady states. The steady states occur when the equilibrium of travel time has been reached, and the network states, such as speed and flow, are stabilized at specific values. \citep{paipuri-2019} included a modified OD matrix as a scenario, while all discussion is based on time-independent TDD conditions. 

Most current model comparison studies have two main limitations: (1) the assumption of uniform trip distance or NE distribution and (2) the static TDD assumption. These assumptions make the input used for models less realistic. The empirical trip distance data have been found to be neither static over the day \cite{Paipuri2020} nor supporting the NE assumption \cite{Martinez2021}. From the perspective of regional dynamic traffic assignment, \citep{Batista2019} argued that the trip distance variability is significant. Further research showed the importance of updating TDD according to traffic conditions \cite{Batista2021}, suggesting that including dynamic TDD could improve the accuracy of the models. 
% There has also been discussion that does not fully agree with the criticism of the NE assumption. \citep{laval2023} found that the difference between the trip-based and the accumulation-based model is only significantly affected by TDD during the transition stage, not in the steady state. The author argued that TDD is formed during job finding and home locating rather than as a set of pre-trip decisions by all travelers. This discussion is limited to a single reservoir with static TDD. 

% and may not be valid when regional dynamic traffic assignment is considered.

To compare the performances of models under a given scenario, it is necessary to include TDD as individual trip distances or categories in the trip-based model and compare it with models under dynamic TDD. Previous studies have primarily compared the effects of different mathematical formulations. An insufficiently studied direction is to use trip distance data with less aggregated TDD information in the models and compare the added value of including TDD in different aggregation levels in the simulation input. 

This paper aims to address the gap by explaining the differences between bathtub models at various TDD aggregation levels and the impact of dynamic TDD on these differences. Demand profiles with different demand-varying speeds are used in the scenarios, helping to understand the impact of demand-varying speeds on model performance. We use two networks based on the road network of Delft and a toy network with a simplified structure, aiming to explore how network properties may affect the performance of bathtub models. The bathtub models are compared with the macroscopic traffic simulations with dynamic traffic assignment. In this way, the differences between bathtub models can be evaluated using the same benchmark. 

The contribution of this paper is threefold. First, we explore the comparison between the accumulation-based model, the M model, and the trip-based models under static and dynamic TDD scenarios in three network cases. Second, we investigate how dynamic TDD conditions and aggregation levels of the TDD might affect the accuracy of bathtub models compared to macroscopic traffic simulations. Third, we examine the influence of network state transition speed on the performance of bathtub models, emphasizing the importance of capturing this network property in bathtub models.
% It is found that the hysteresis loop may appear on the free-flow branch of the MFD when the network has less or no alternative routes. This should be noticed in networks with low connectivity, especially when multiple road hierarchies are included. 

The rest of this paper is organized as follows: in the next section, bathtub models, MFD function and the networks used in the simulation are described. The third section introduces data inputs, MFD estimation results, and the simulation scenarios. The fourth section presents the simulation results. Conclusions and discussions are given in the fifth section. 

\section{Methodology}
\subsection{Bathtub models}
The accumulation-based model \cite{Vickrey2020, Daganzo2007}, the M-model \cite{Lamotte2018, Sirmatel2021}, and two trip-based simulation methods with different time discretizations are used in this paper. The main conservation equation in bathtub models is Eq.\ref{eq:daganzo accumulation}, where $n(t)$ is the network accumulation, $e(t)$ and $g(n(t))$ are the inflow and outflow at time $t$, respectively. In the accumulation-based model, the expression of outflow is Eq.\ref{eq:daganzo outflow}, where $V(n(t))$ is the speed-accumulation relationship represented by the MFD, $D$ is the MTD of the network. \\ 
\begin{equation}
    \Dot{n}(t) = e(t)-g(n(t))
    \label{eq:daganzo accumulation}
\end{equation}
\begin{equation}
    g(n(t)) = n(t)V(n(t))/D
    \label{eq:daganzo outflow}
\end{equation} \\
The M-model is derived from the conservation of the total remaining trip distance $m(t)$, so that the past of the system can be included in the state function of the network. As shown in Eq.\ref{eq:M accumulation}, the change in $m(t)$ is defined as the difference between the added trip distance $e(t)L$ and the distances traveled $n(t)V(n(t))$ during time step $t$. The outflow expression for the M-model is shown in Eq.\ref{eq:M outflow}. The added term $\alpha(m(t)/n(t)D^*-1)$ is used to adjust the outflow when $m(t)$ is not equal to the total remaining distance in the steady state. The latter is represented by the network accumulation and a steady trip distance $D^*$, which differs among different TDD types. $D^*$ is calculated using the mean $D$ and variation $\sigma^2$ of the TDD, as shown in Eq.\ref{eq:M steady distance}. The outflow will be lower when $m(t)$ is higher than the remaining distance in steady states and higher when $m(t)$ is lower than the remaining distance in steady states. When the TDD follows NE distribution, the trip distance at steady state $D^*$ is equal to the mean $D$, as $D = \sigma$ for all NE distributions. In this case, the M-model is equivalent to the accumulation-based model. \\
\begin{equation}
    \Dot{m}(t) = e(t)D - n(t)V(n(t))
    \label{eq:M accumulation}
\end{equation}

\begin{equation}
    g(n(t)) = \frac{n(t)V(n(t))}{D} \left( 1+\alpha \left(\frac{m(t)}{n(t)D^*}-1\right) \right)
    \label{eq:M outflow}
\end{equation}

\begin{equation}
    D^* = \frac{D^2+\sigma^2}{2D}
    \label{eq:M steady distance}
\end{equation} \\
There is no general rule for selecting the value of $\alpha$. \citep{Lamotte2018} derived $\alpha = -3$ when the TDD follows a gamma distribution with specific parameters. \citep{Sirmatel2021} used the optimized $\alpha$ calibrated by minimizing the error between M-model prediction and the data. In the present paper, $\alpha$ is also optimized to minimize the RMSE between simulation results with macroscopic traffic simulation data. This ex post strategy makes it impossible for M-model to have time-varying $\alpha$ to adapt to dynamic TDD.

In the trip-based simulation, trips are considered individually. Eq.\ref{eq:Tripbased length} shows that the trip distance of a trip $i$ is the integration of the speed of the network when trip $i$ is inside the network. $t-T_i(t)$ and $t$ are respectively the entry time and exit time of trip $i$. \\
\begin{equation}
    D_i = \int_{t-T_i(t)}^t V(n(s))ds
    \label{eq:Tripbased length}
\end{equation}\\
Trip-based simulation can be built with different time discretizations. An event-based simulation method based on the trip-based model was proposed by \citep{Mariotte2017}, where the system states are updated whenever a trip enters or exits the network. It is a suitable tool to handle variety in trip distances. With fixed time steps, \citep{Mariotte2017} built a trip-based simulation with homogeneous trip distance assumption. Agent-based Bathtub Model (AB2M) is proposed by \citep{martinez-2024}. It is another trip-based model using fixed time steps, where a characteristic trip distance \cite{jin-2020} is included to handle the entries and leaves of trips more efficiently. When the length of time steps is fixed, trips are assumed to have the same speed during each time step. In this study, event-based simulation and trip-based model with fixed time steps will both be included for model comparison. 

\subsection{MFD estimation method}
A function form of MFD introduced by \citep{ambuhl-2020} is used to fit the speed accumulation data points obtained from the macroscopic traffic simulation. This function form showed better accuracy compared to other MFD estimation methods in four different cities \cite{ambuhl-2020}. As shown in Eq.\ref{eq:function mfd}, the original function is an average flow expression (referred to in other studies as the p-MFD). $q(k)$ is the average flow under density $k$, $u_f$ is the free-flow speed of the network, Q is the capacity average flow of the network, $\kappa$ is the jam density of the network, and $w$ is the wave speed of the network. $\lambda$ is a smooth parameter with a suggested range between 0.03 to 0.07 based on empirical studies in \cite{ambuhl-2020}. With these parameters available, this function form MFD can provide accurate estimations without requiring a considerable amount of speed-accumulation data \cite{ambuhl-2020}. This makes this function form suitable for this study, as the demand profiles are defined with fictitious data and cannot generate a considerable amount of representative data. In other words, this function form MFD can include more network properties and narrow the range of the MFD fitting parameters, compensating for the lack of speed-accumulation data in this study.\\
\begin{equation}
    q(k) = -\lambda ln\left( exp\left( \frac{u_fk}{\lambda}\right)+exp\left(\frac{Q}{\lambda}\right)+exp\left(\frac{-(\kappa-k)w}{\lambda} \right) \right)
    \label{eq:function mfd}
\end{equation}\\
The parameters related to network properties were first calculated ($u_f$ and $Q$) or assigned with typical values found in empirical studies ($\kappa$ and $w$) \cite{ambuhl-2020}. The final MFD parameters used in simulations were calibrated through non-linear least squares, which minimizes the RMSE of MFD curves compared to speed-accumulation data points. The range of network property parameters is defined as $\textbf{X} \pm 0.2\textbf{X}$, where \textbf{X} represents the initial values of these parameters. The smoothing parameter $\lambda$ can change within the suggested range by \citep{ambuhl-2020}, with an initial value of 0.03.

% The expression of the error in total trip production is shown in Eq \ref{eq:error prod}. Given a time resolution $i$, its time step length is $dt_i$. $N_i$ represents the total number of vehicles calculated for a time step. $L_j$ represents trip distance of trip $j$. $L$ and $q$ are the mean trip distance and inflow in the macrosimulation. The error in production $E^{production}_i$ is normalized by $dt_i$ and has a unit of veh*km/h. For the AB2M and event-based simulation using the me n trip distance, $L_j = L$. This means that minimizing the error in production is equivalent to minimizing the error in trip numbers, resulting in the expression in Eq \ref{eq:error vnr}. The optimal time step length for AB2M simulations is the one that has the lowest total numerical error under all demand levels. 
% \begin{equation}
%     E^{production}_i = \frac{\sum_{j = 1}^{N_i} L_j - L*q*dt_i}{dt_i}
%     \label{eq:error prod}
% \end{equation}
% % \begin{equation}
% %     E^{distance}_i = \frac{\sum_{j = 1}^{N_i} L_j}{N_i} - L 
% %     \label{eq:error mean}
% % \end{equation}
% \begin{equation}
%     E^{trip number}_i = \frac{N_i - q*dt_i}{dt_i}
%     \label{eq:error vnr}
% \end{equation}

\subsection{Network description}
This paper considers the road network around the city of Delft in the Netherlands. The city covers an area of 24 $km^2$. Delft's morning peak traffic was modeled by using the macroscopic traffic simulation tool OmniTrans \cite{goudappel}. The socio-demographic data of the traffic analysis zones are fictitious and are used for educational usage. Although the input and output of the simulation cannot represent the real traffic in Delft, it can still be a reference case in this study, as the primary purpose is to compare the models' performance. 
\begin{figure}[h!]
     \centering
     \begin{subfigure}[b]{0.4\textwidth}
         \centering
         \includegraphics[width=\textwidth]{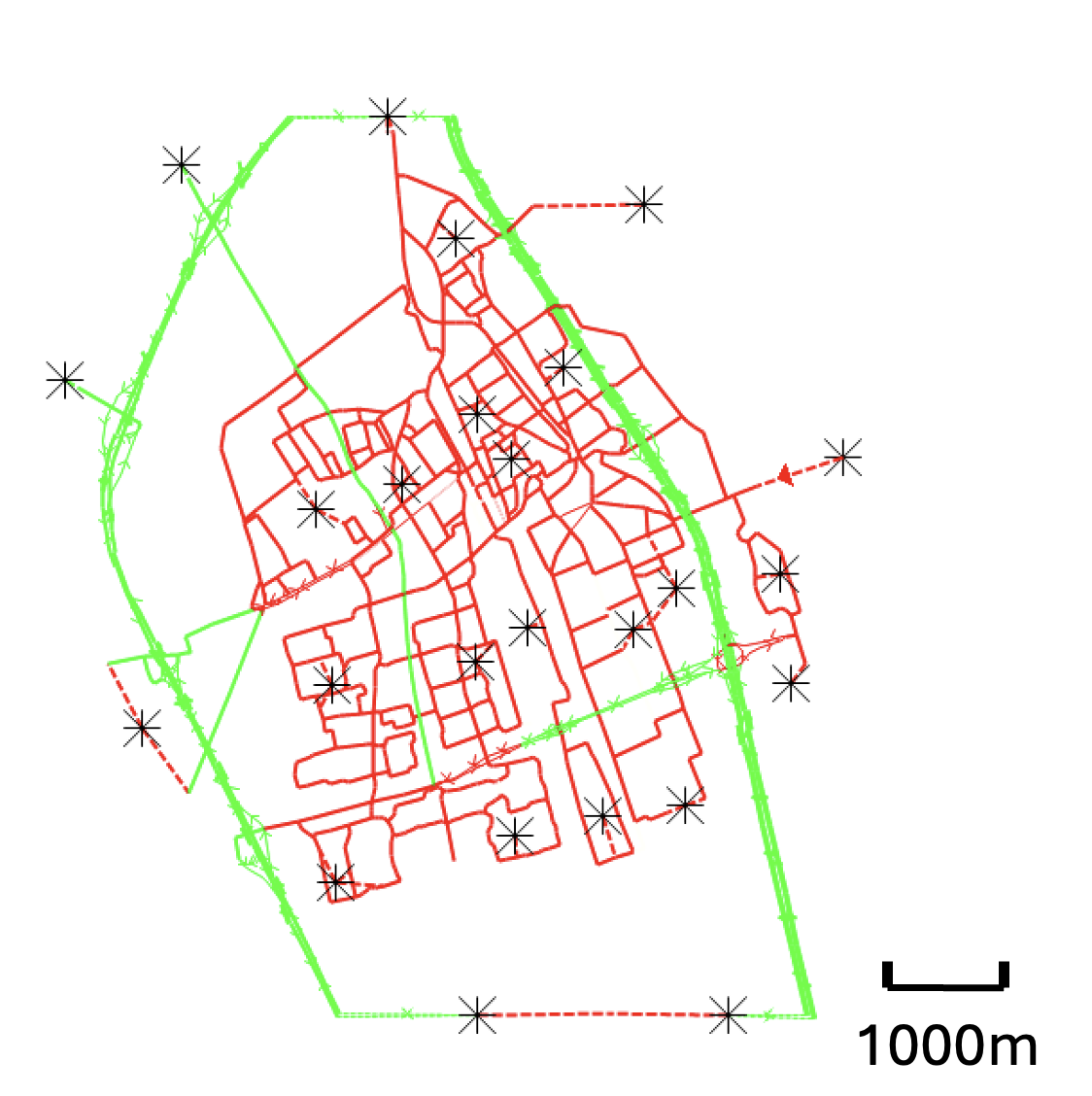}
         \caption{}
         \label{fig:network delft}
     \end{subfigure}
     \hfill
     \begin{subfigure}[b]{0.4\textwidth}
         \centering
         \includegraphics[width=\textwidth]{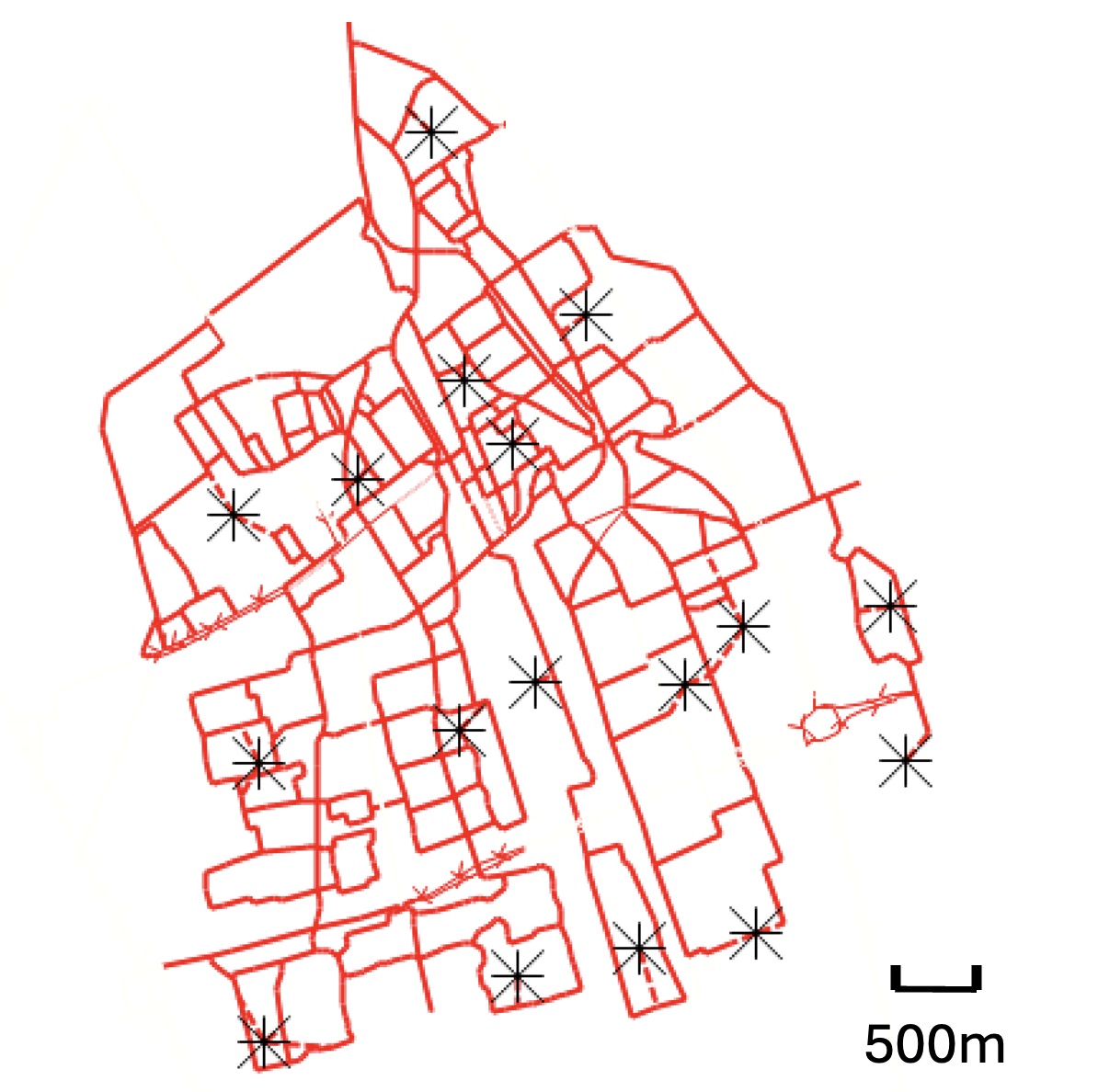}
         \caption{}
         \label{fig: network delft int}
     \end{subfigure}

    \vspace{0.5cm}
    \begin{subfigure}[b]{0.62\textwidth}
         \centering
         \includegraphics[width=\textwidth]{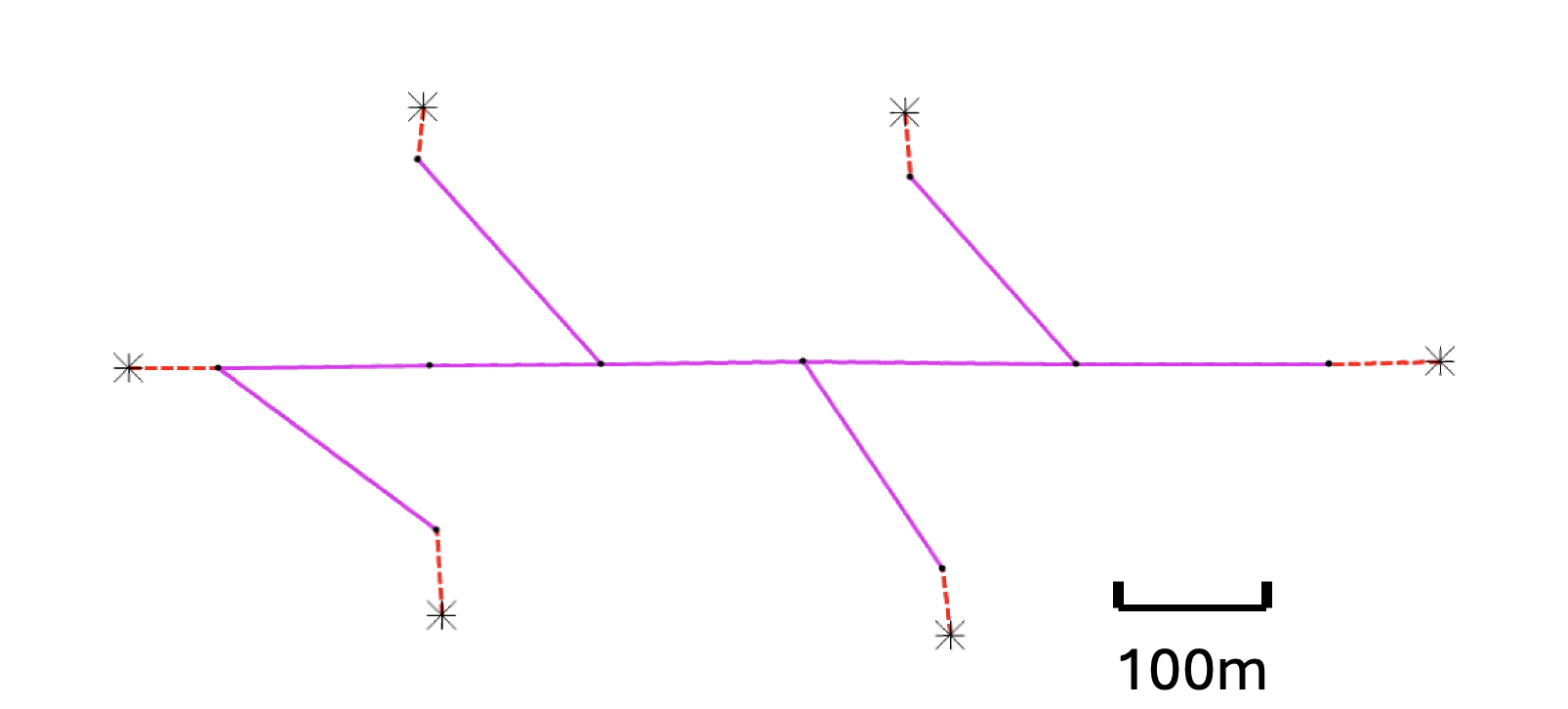}
         \caption{}
         \label{fig:network_simp}
     \end{subfigure}
        \caption{Road networks used for macroscopic traffic simulation. Stars represent centroids of zones and the connectors are represented by dashed links. (a) Delft network with freeways (DF); (b) Delft urban network (DU); (c) Toy network (T).}
        \label{fig:networks}
\end{figure}

The network of Delft with freeways (DF) is shown in Figure \ref{fig:network delft}. There are over 500 links with speed limits ranging from 35km/h to 100 km/h, with a total lane distance of 213km$\cdot$lane. Green and red links represent urban roads (speed limit lower than 70km/h) and freeways (speed limit higher than or equal to 70km/h), respectively. 7 external zones and 18 internal zones are considered in the network. Traffic from (and to) external zones represents the demand from (and to) adjacent cities (Den Hoorn, Rijswijk, Den Haag, and Rotterdam), while traffic from and to internal zones represents the demand in Delft. 

To explore the influence of freeways, the urban road network of Delft (DU) with a higher speed homogeneity is used as a separate case. The structure of the urban network is shown in Figure \ref{fig: network delft int}. Only the 18 internal zones are included in this case. The total lane distance of this network is 94km$\cdot$lane. 
% The properties of this network make it atypical among MFD-based traffic simulation studies, as there are more road hierarchies, including freeways. This also makes the connectivity of this network not as high as a district in a metropolitan, as local congestion happens because no alternative route with close travel time costs is available, leading to less re-routing behavior. This makes the network suitable for exploring the effect of dynamic TDD caused by time-dependent OD matrices. 

A toy network (T) is created to better understand the effect of multiple road hierarchies. As shown in Figure \ref{fig:network_simp}, this network has only 6 zones and 14 links. This network contains links with the same speed limit of 35km/h. To reduce the numerical error caused during generating individual trips, more trips are needed. To avoid bottleneck congestion in the toy network, the network size and demand are both upscaled 10 times, resulting in 20 lanes per link and a total lane distance of 58km$\cdot$lane. The assumption of 20 lanes aims to create a homogeneous and well-connected network in contrast with the two Delft networks. In the toy network, each OD pair has only one possible route. This means that trip distances for each OD pair will not be influenced by re-routing, and the TDD of the network is static, given a static OD matrix.

\section{Simulation setting}
\subsection{Input data}
In OmniTrans, the dynamic traffic assignment uses the model MaDAM \cite{goudappel}, which uses the paired combinatorial logit model (PCL) in the dynamic traffic assignment. \citep{Dijkhuis2012} discussed that the deterministic dynamic user equilibrium (DUE) cannot be reached using PCL in OmniTrans, as PCL always assigns traffic to every available route. The iteration ends based on the iteration number limit or duality gap between route costs. This can not guarantee the same travel time among all routes of an OD pair. This drawback of MaDAM has limited influence on the total accumulation of the network but may have a larger effect on TDD when the network is congested and there is more re-routing behavior. 

The link-level macroscopic traffic simulation data are aggregated to the accumulation and average speed of the network. The aggregation is done using Eq.\ref{eq:accumulation aggregation} and \ref{eq:speed aggregation}. The set of links is represented by $E$. The accumulation at a time step $n$ is the sum of the vehicle numbers on every link $l$ in the network. The vehicle number can be calculated with the density $k_l$ [veh/km] and the lane distance of the link $d_l$ [km$\cdot$lane]. The average speed of the network is the weighted average of link speed $v_l$ with the number of vehicles as the weight. 
In the two Delft networks, some links have very low density during the simulation. These links are not counted in the total network lane distance $L_N$. The calculation of $L_N$ is shown in Eq.\ref{eq:total lanekm}. Let $E_f$ be the subset of $E$, in which links have densities not lower than a certain threshold. In all three networks, the threshold is selected as 3veh/km. \\
\begin{equation}
    n = \sum_{l \in E} k_ld_l
    \label{eq:accumulation aggregation}
\end{equation}
\begin{equation}
    v = \frac{\sum_{l \in E} v_lk_ld_l}{n}
    \label{eq:speed aggregation}
\end{equation}
\begin{equation}
    L_N = \sum_{l \in E_f} d_l
    \label{eq:total lanekm}
\end{equation} \\
With the demand for each OD and the traffic assignment results, the TDD can be obtained. Due to the continuous demand in macroscopic traffic simulation, it is not possible to include every trip individually in the TDD. Therefore, TDD can only be derived with aggregated trip distance categories. As shown in Eq.\ref{eq:proportion}, the proportion of trips with a trip distance in category $d$ is represented by $p_d$. $R_d$ is the set of all routes with a length in category $d$. $N$ is the set of all traffic analysis zones. Proportion $p_{jkr}$ represents the traffic assignment result. $q_{jk}$ is the total flow from origin $j$ to destination $k$. $p_{jkr}q_{jk}$ is the assigned flow on route $r$ from origin $j$ to destination $k$. 
The MTD for each trip distance category $D_d$ is calculated using Eq.\ref{eq:mean distance}, where $d_r$ is the length of route $r$. \\
\begin{equation}
    p_d = \frac{\sum_{r \in R_d}\sum_{j \in N} \sum_{k \in N} p_{jkr}q_{jk}}{\sum_{j \in N} \sum_{k \in N} q_{jk}}
    \label{eq:proportion}
\end{equation}

\begin{equation}
    D_d = \frac{\sum_{r \in R_d}\sum_{j \in N} \sum_{k \in N} d_rp_{jkr}q_{jk}}{\sum_{j \in N} \sum_{k \in N} q_{jk}}
    \label{eq:mean distance}
\end{equation} \\
For the two Delft networks, it is assumed that the TDD is time-independent when the internal structure of the OD matrix is not changed. The TDD of the whole simulation can be represented by the TDD derived from one time step in the macroscopic traffic simulation. The assumption on time-independent TDD implies that the possible re-routing behavior in dynamic traffic assignment will not influence the TDD considerably.

For the trip-based models, two levels of TDD aggregation are used as the input to explore the effect of using TDD at different aggregation levels for trip-based models. The first scenario uses only the MTDs, leading to a homogeneous trip distance for each individual trip. The second scenario uses a discrete distribution that contains several trip distance categories. In microsimulations, the exact trip number and deterministic distance of each trip can be obtained and input to the trip-based simulations as individuals. To create individual trip distances for the trip-based models, the number of trips with certain trip distances should be calculated using the proportion for each time step. For trip-based models, the trip numbers are calculated per 60s, which is also the time resolution in the macroscopic traffic simulation. This ensures that the whole TDD can be represented in every 60s. The error in inflow can be caused by rounding vehicle numbers to integers in trip-based models. With a demand level of around 20000 veh/h in two Delft networks and the toy network, 60s is large enough not to cause influential numerical errors. With a time step of 60s, the errors are less than 0.2\% in the two Delft networks and the toy network case. 

% \begin{equation}
%     E^{trip number}_i = \frac{N_i - q*dt_i}{dt_i}
%     \label{eq:error vnr}
% \end{equation}

For the trip-based model with fixed time steps, states are updated each time step, which assumes the speed to be homogeneous among all trips within each step. This assumption can lead to significant differences in simulation results when the time step is too long. To ensure that the selected time aggregation level has limited influence on the performance of the trip-based model with fixed time steps, convergence tests were conducted in all scenarios. For each case, the time resolution is reduced in each test. If the mean difference in results from two consequential tests is less than 1\%, the model is considered converged. A time step of 2s is used for the DF network and the DU network, and 0.5s is used for the toy network. 

\subsection{MFD calibration}

\begin{figure}
     \centering
     \begin{subfigure}[b]{0.48\textwidth}
         \centering
         \includegraphics[width=\textwidth]{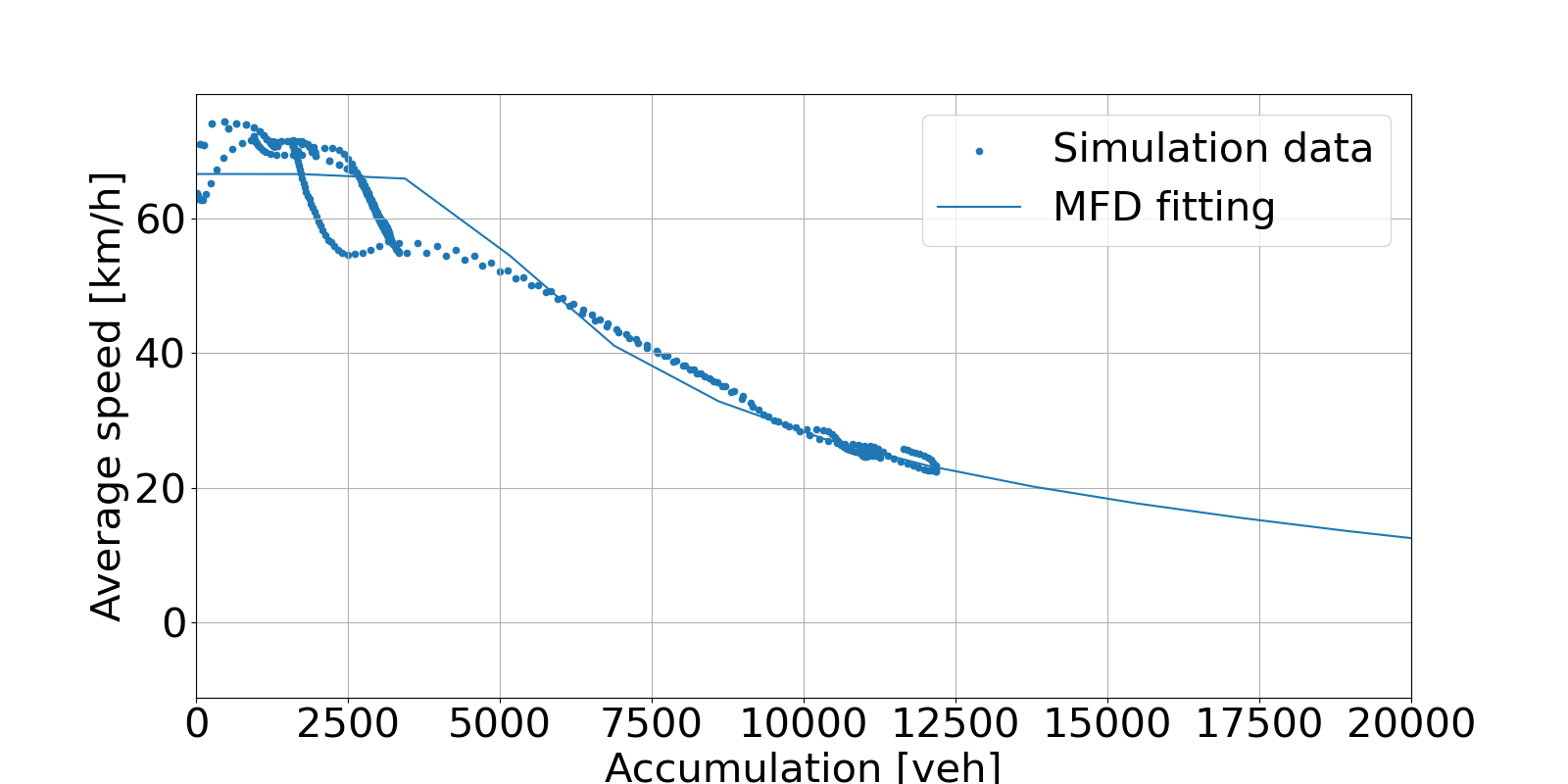}
         \caption{}
         \label{fig:mfddelft}
     \end{subfigure}
     \hfill
     \begin{subfigure}[b]{0.48\textwidth}
         \centering
         \includegraphics[width=\textwidth]{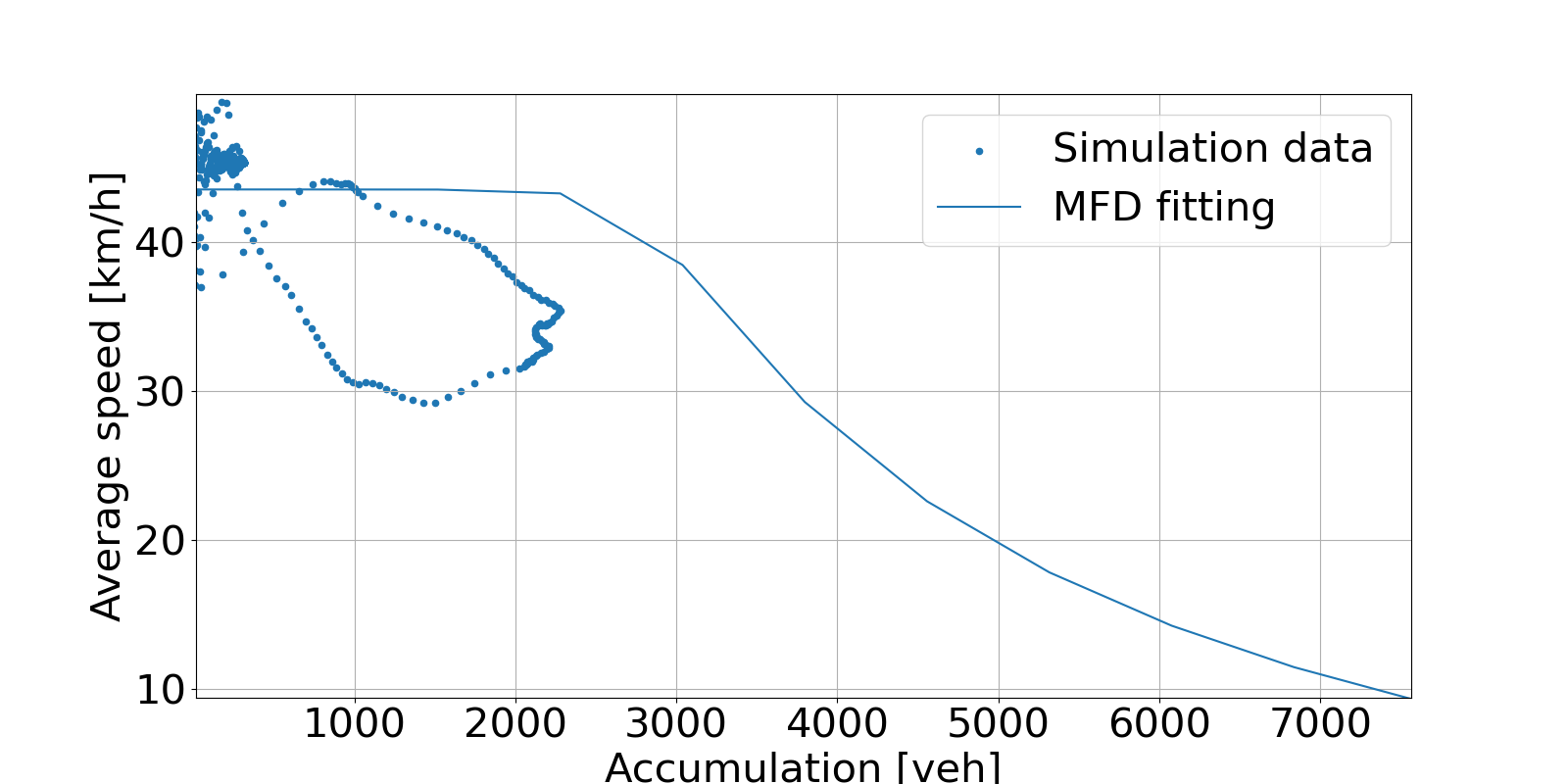}
         \caption{}
         \label{fig:mfddelftint}
     \end{subfigure}
     \vspace{0.5cm}
     \begin{subfigure}[b]{0.48\textwidth}
         \centering
         \includegraphics[width=\textwidth]{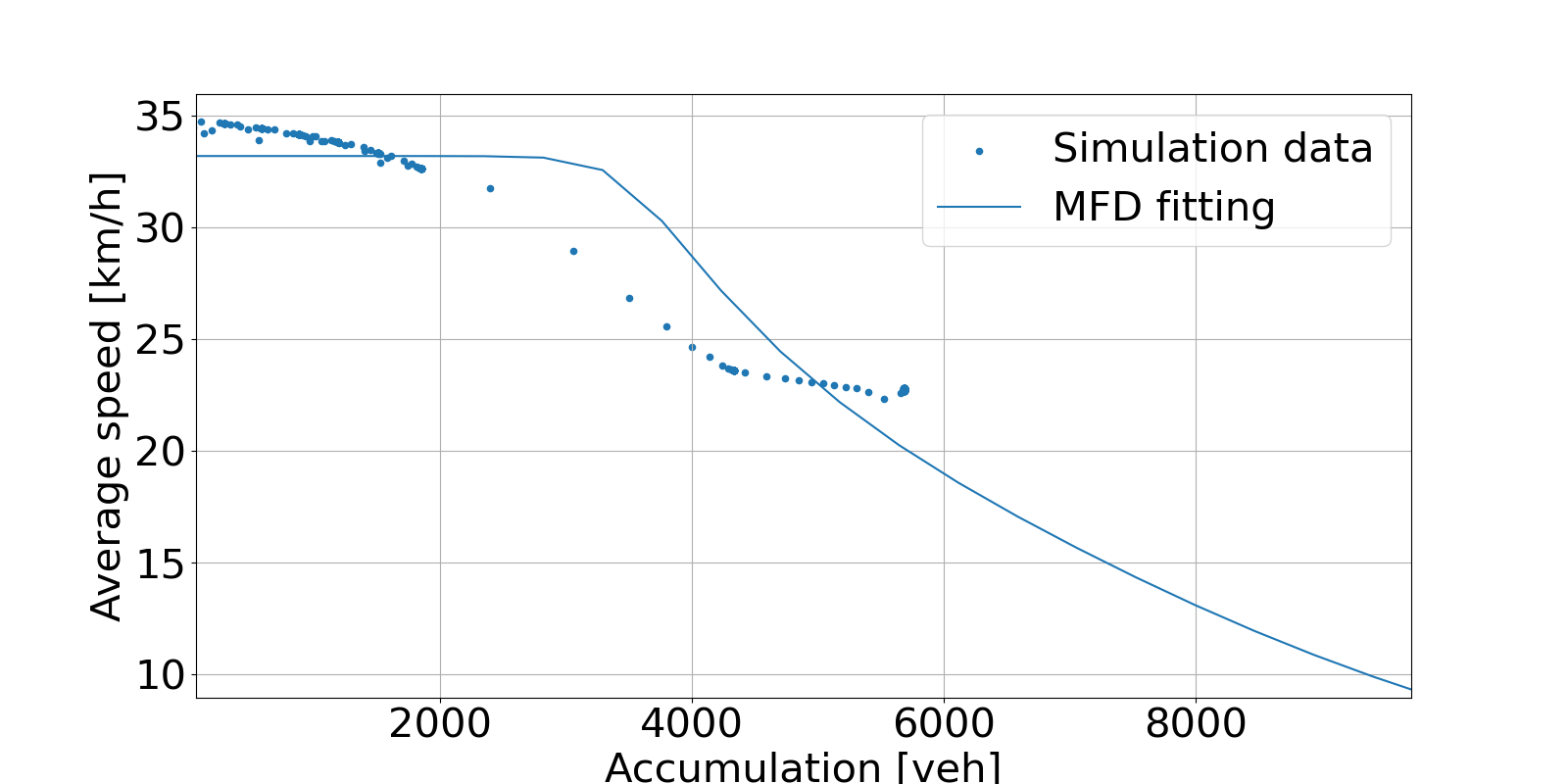}
         \caption{}
         \label{fig:mfdsimp}
     \end{subfigure}
        \caption{Speed MFD fitting results for three cases. (a) Delft network with freeways (DF); (b) Delft urban network (DU); (c) Toy network (T).}
        \label{fig:mfd}
\end{figure}

Figure \ref{fig:mfd} shows the fitted MFD curve and the collected data points in the two Delft networks and the toy network. Data from uncongested and congested situations are used in the MFD estimation. For each MFD estimation, data is derived from two demand scenarios that result in no congestion in the network and one demand scenario that causes severe congestion in the network. Demand profiles used in these scenarios are tuned in the macroscopic simulation, with the goal of creating free-flow and congested scenarios. The estimated parameters ($\lambda$, $u_f$[m/s], $Q$[veh/s], $\kappa$[veh/m], $w$[m/s]) for the DF network are: (0.034, 19.2, 0.18, 0.43, 2.42). For the DU network, the estimated parameters are: (0.03, 12.1, 0.15, 0.57, 3). In the toy network, the estimated parameters are: (0.03, 9.2, 0.34, 0.55, 2.5).

It can be seen in Figure \ref{fig:mfd} that the estimated MFD curves are less accurate at the points when the speed starts reducing. This is related to the lack of alternative routes in the networks. The differences between estimated MFD curves and data points suggest that the congestion starts at an accumulation level lower than the critical accumulation estimated in the MFD. The speed reduction starts earlier than predicted because congestion grows locally instead of spreading evenly on the network. For the DF network, a hysteresis loop is observed on the free-flow branch of the speed MFD. This will be further explained in the simulation result section. The only case without data on the congestion branch is the DU network (Figure \ref{fig:mfddelftint}). This may be because the DU network has the highest connectivity among the three networks. Thus, local congestion can spread more uniformly to the whole network.

\begin{figure}[htbp]
    \centering
    \begin{subfigure}[b]{0.48\textwidth}
        \centering
        \includegraphics[width=\textwidth]{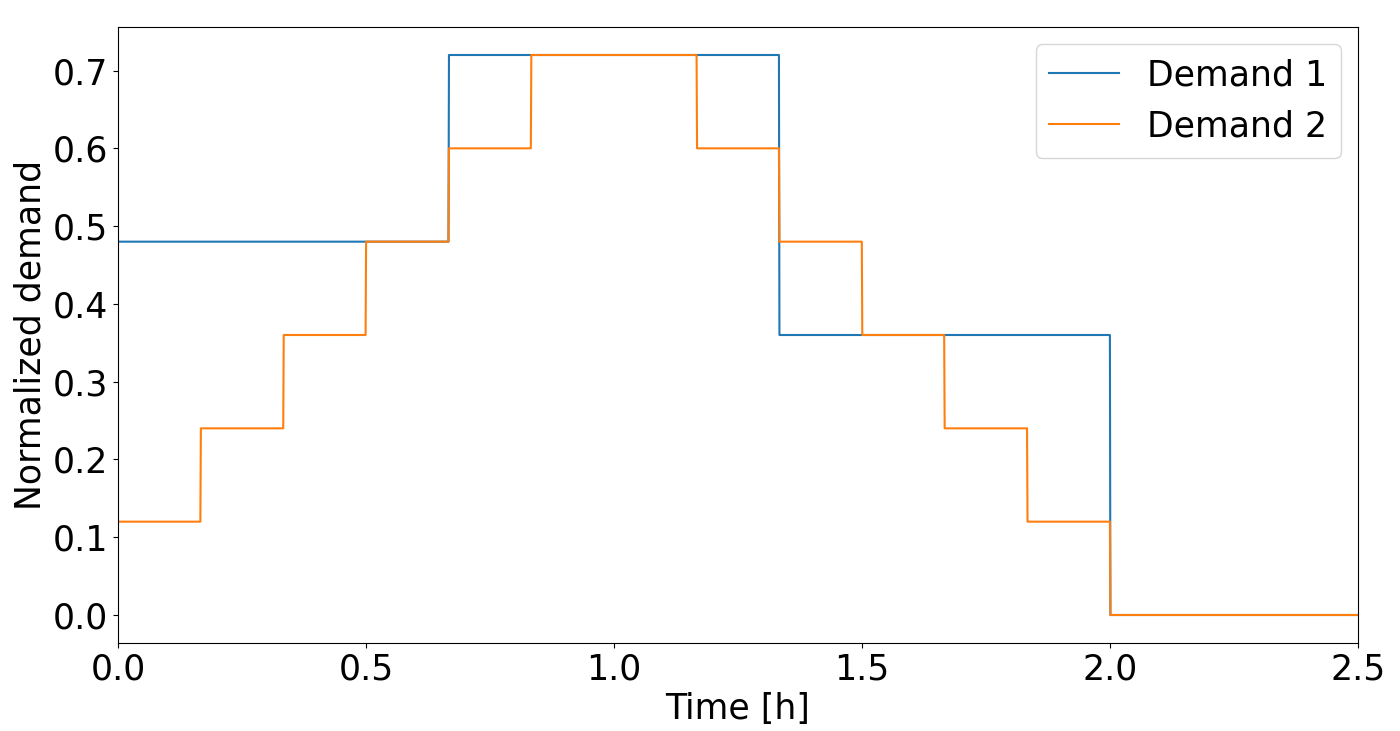}
        \caption{}
        \label{fig:demand curves}
    \end{subfigure}
    \hfill
    \begin{subfigure}[b]{0.48\textwidth}
        \centering
        \includegraphics[width=\textwidth]{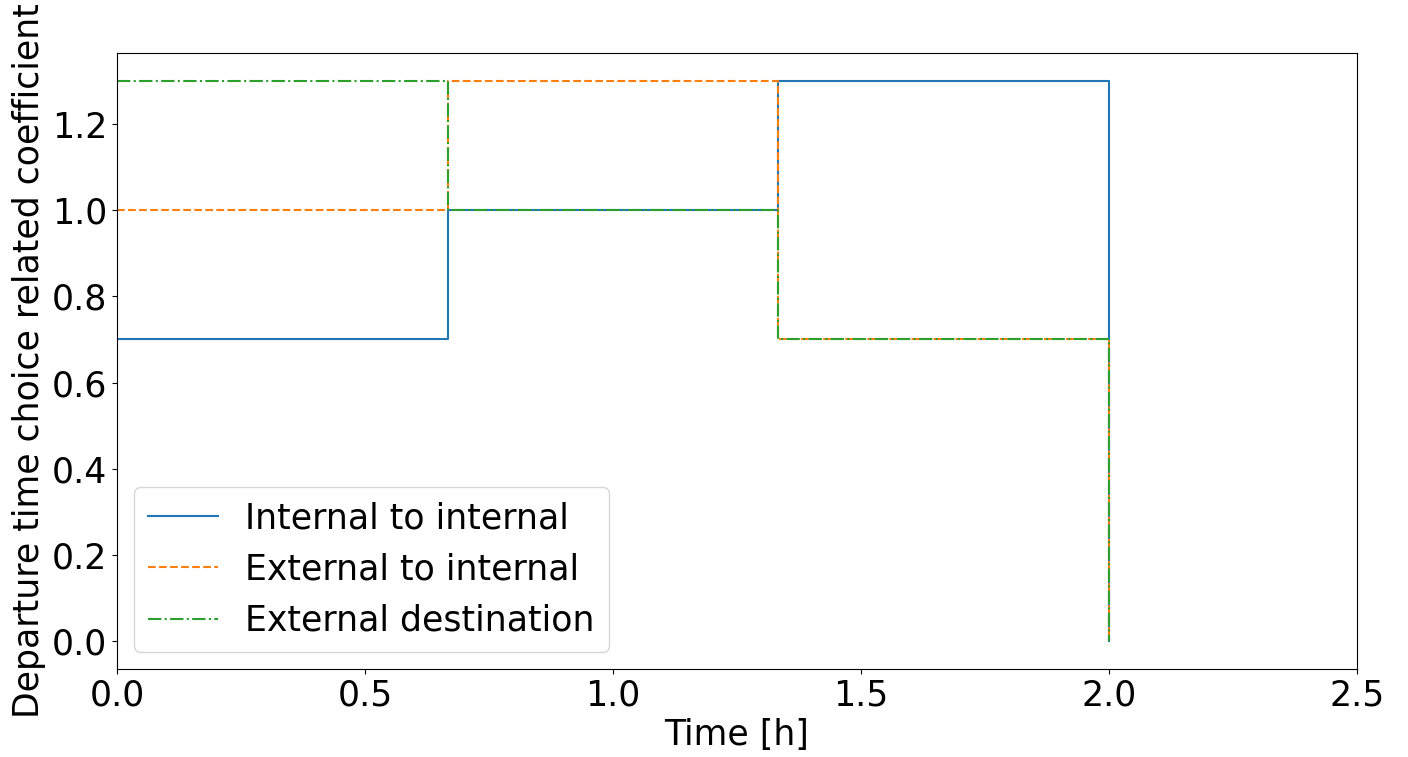}
        \caption{}
        \label{fig:coefficients}
    \end{subfigure}
    \vspace{0.5cm}
    \begin{subfigure}[b]{0.48\textwidth}
        \centering
        \includegraphics[width=\textwidth]{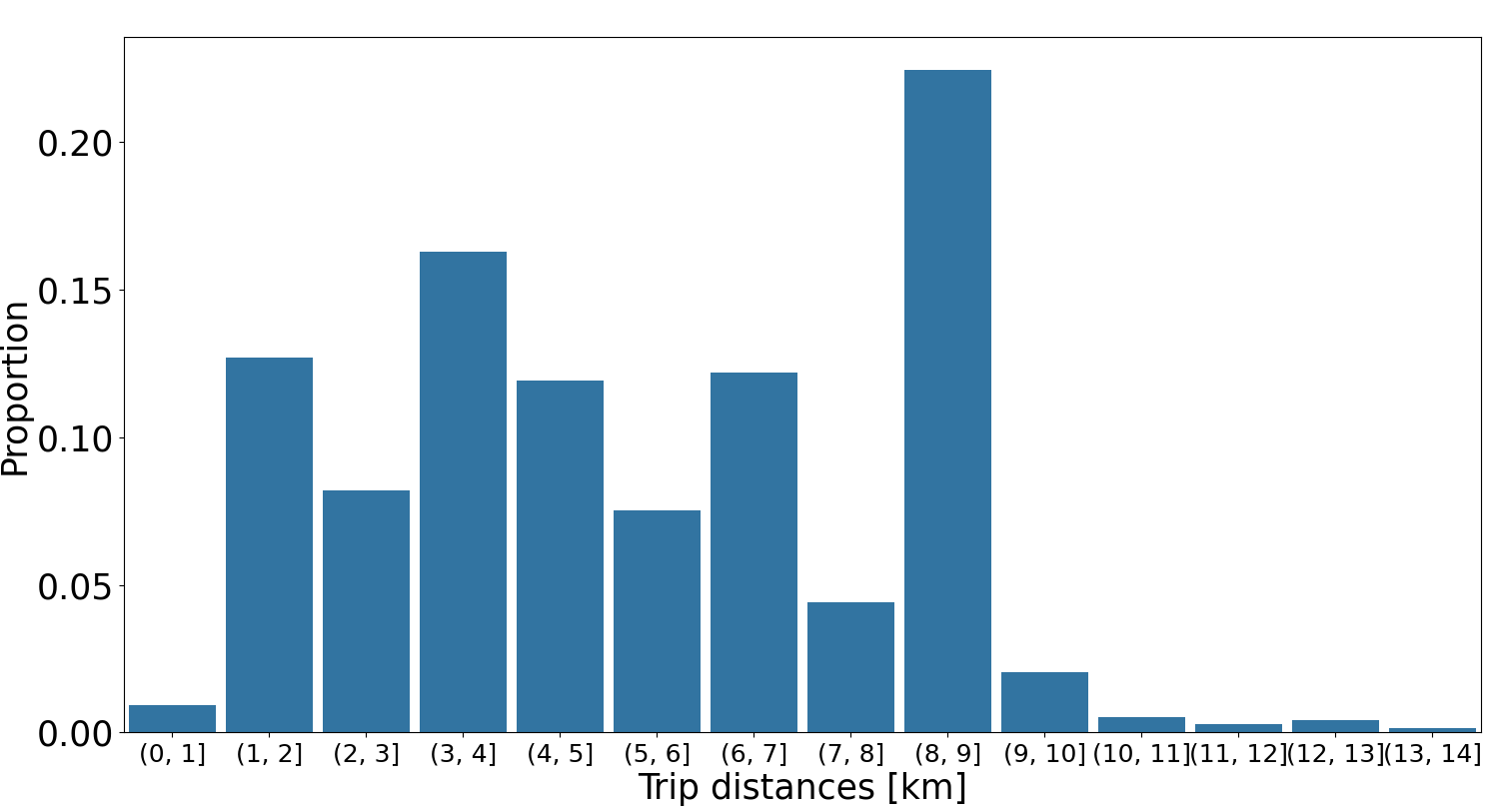}
        \caption{}
        \label{fig:del_TDDsta}
    \end{subfigure}
    \hfill
    \begin{subfigure}[b]{0.48\textwidth}
        \centering
        \includegraphics[width=\textwidth]{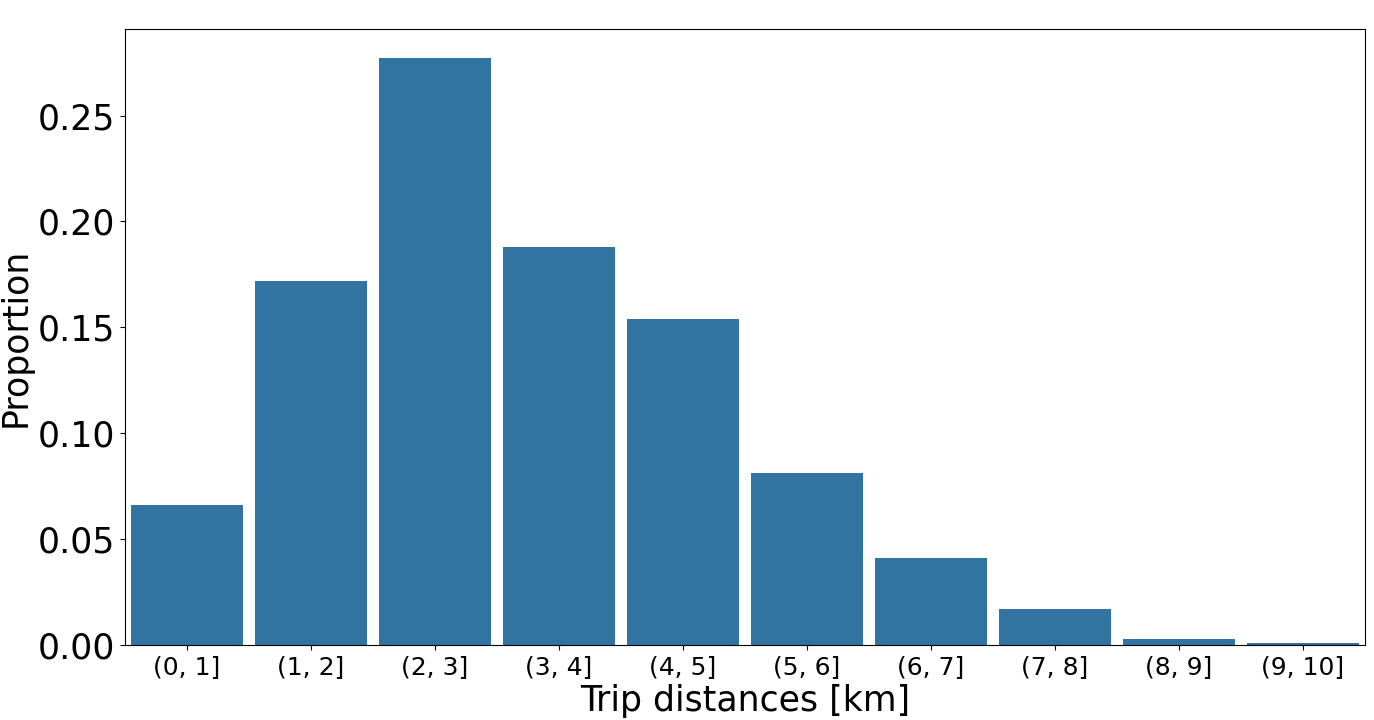}
        \caption{}
        \label{fig:int_TDDsta}
    \end{subfigure}
    \vspace{0.5cm}
    \begin{subfigure}[b]{0.48\textwidth}
        \centering
        \includegraphics[width=\textwidth]{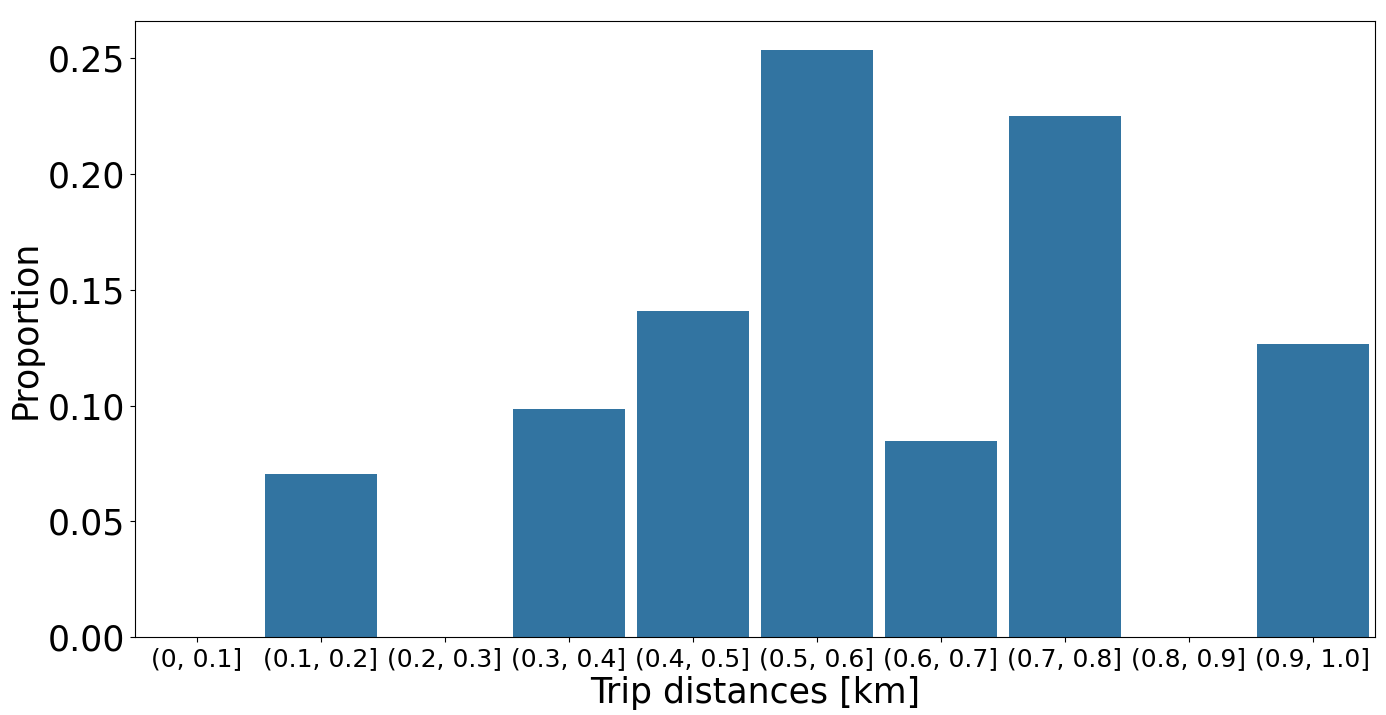}
        \caption{}
        \label{fig:simp_TDDsta}
    \end{subfigure}
    \hfill
    \begin{subfigure}[b]{0.48\textwidth}
        \centering
        \includegraphics[width=\textwidth]{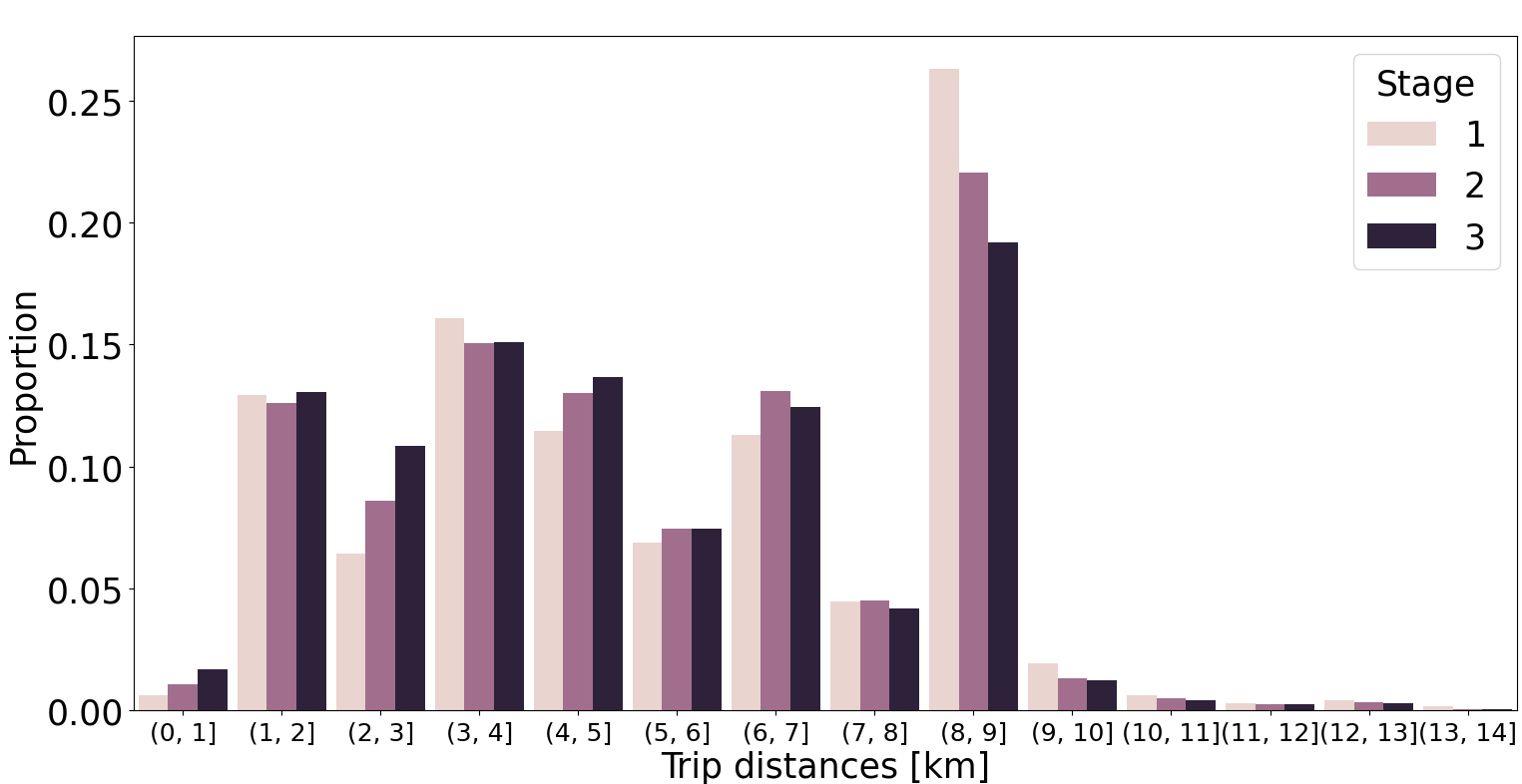}
        \caption{}
        \label{fig:dynatdd}
    \end{subfigure}
    \vspace{0.5cm}
    \begin{subfigure}[b]{0.58\textwidth}
        \centering
        \includegraphics[width=\textwidth]{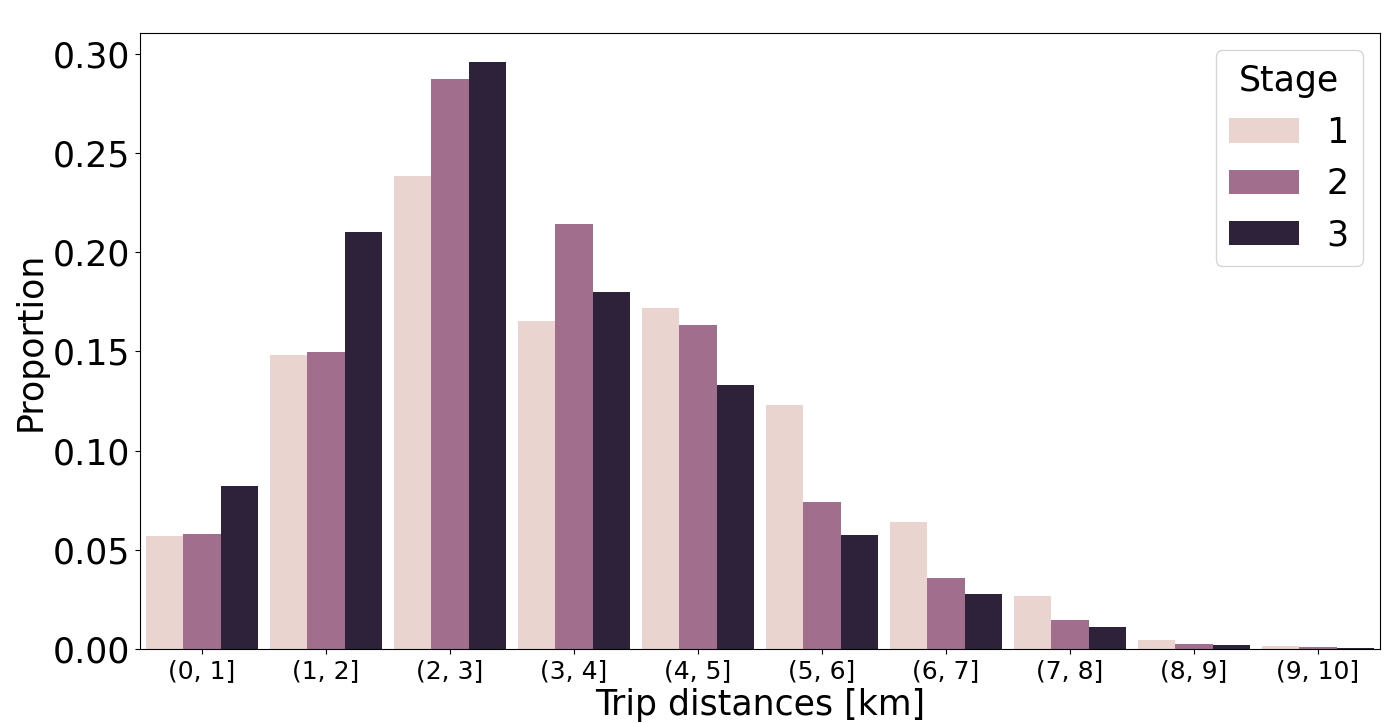}
        \caption{}
        \label{fig:dynatdd_int}
    \end{subfigure}

    \caption{Conditions in different simulation scenarios. (a) Two demand profiles; (b) Coefficients for time-dependent OD matrices; (c) Delft network with freeways (DF): static TDD; (d) Delft urban network (DU): static TDD; (e) Toy network (T): static TDD; (f) Delft network with freeways (DF): dynamic TDDs; (g) Delft urban network (DU): dynamic TDDs.}
    \label{fig:main}
\end{figure}
\subsection{Simulation scenarios}
As shown in Figure \ref{fig:demand curves}, two demand profiles are used in the three networks. Demand profile 1 has more steep demand changes, while demand profile 2 has a more smoothed peak. These two profiles are used to test the difference in model performance under fast and slow varying demand conditions. Although re-routing is an essential factor affecting TDD, the main focus is dynamic TDD caused by the time-dependent OD matrices in this research. To reduce the effect of re-routing in three networks, the applied demand profiles are tuned not to lead to congested network states. 

In dynamic TDD scenarios in the DF network, the OD matrix is considered to be time-dependent. Departure time choice is assumed to be the same among all trips when the same demand profile is applied. In the dynamic TDD scenarios, assumptions were made for trips with different kinds of ODs to create dynamic OD matrices. The coefficients used for time-dependent OD matrices are shown in Figure \ref{fig:coefficients}. The assumption is that every trip has the same desired arrival time. Thus, longer trips will prefer to depart earlier than shorter ones. For trips traveling from internal zones to internal zones, the peak of departure time distribution is assumed to be at the end of the morning peak period. For trips traveling from external zones to internal zones, the peak is assumed to be in the middle of the morning peak period. For all trips with external destinations, the peak is assumed to be at the beginning of the morning peak period. These assumptions aim to create considerable differences in the OD matrix and, hence, the TDD. They are not representative enough to reflect departure time distribution in real life. For the dynamic TDD case in the DU network, demand between OD pairs is categorized into three categories based on trip distances. The dynamic TDDs are also created using the coefficient in Figure \ref{fig:coefficients}.  

The TDD in the DF network is shown in Figure \ref{fig:del_TDDsta}; the demand is distributed between 0km and 13km with a MTD of 5.28km. The high proportion of trip distances of 8km is due to the high demand from external zones to external zones. In the DU network, trips are distributed between 0km and 9km (as shown in Figure \ref{fig:int_TDDsta}). The MTD is 3.23km. The TDD of the toy network (Figure \ref{fig:simp_TDDsta}) has three bars with no demand. This is because there are limited route lengths in this simplified network layout. 
Chi-square tests were performed on the TDDs derived from different time points in the macroscopic traffic simulation. All TDDs are not significantly different from each other. This supports the assumption of time-independent TDD when there is no internal change in the OD matrix. 

Figure \ref{fig:dynatdd} shows the dynamic TDD used for the DF network. Three stages correspond to the three steps of demand adjusting coefficients in Figure \ref{fig:coefficients}. The MTDs for each stage are 5.48km, 5.24km, and 5.01km. The dynamic TDDs of the DU network are shown in Figure \ref{fig:dynatdd_int}. The MTDs for each stage are 3.55km, 3.25km, 2.95km. 
\begin{table}[htbp]
\centering
\captionsetup{justification=centering}
\caption{Overview of simulation scenarios}
\label{tab:scenarios}
\begin{tabular}{llll}
\begin{tabular}[c]{@{}l@{}}Scenario \\ codes\end{tabular} & Network & TDD     & Demand profile \\ \hline
DF-S-1        & Delft network with freeways                        & Static  & 1 (Fast-changing)                                                         \\
DF-S-2        & Delft network with freeways                        & Static  & 2 (Slow-changing)                                                         \\
DU-S-1        & Delft urban network               & Static  & 1                                                         \\
DU-S-2        & Delft urban network               & Static  & 2                                                         \\
T-S-1        & Toy network                  & Static  & 1                                                         \\
T-S-2        & Toy network                  & Static  & 2                                                         \\
DF-D-1        & Delft network with freeways                        & Dynamic & 1                                                         \\
DF-D-2        & Delft network with freeways                        & Dynamic & 2                                                         \\
DU-D-1        & Delft urban network               & Dynamic & 1                                                         \\
DU-D-2       & Delft urban network               & Dynamic & 2                                                         \\ \hline
\end{tabular}
\end{table}

The scenarios are summarized in Table \ref{tab:scenarios}. Simulations under static TDD are done for all three networks with both demand profiles. In the dynamic TDD scenarios, the DF network and the DU network are compared. 

\section{Simulation results}
The OmniTrans simulations were run on a laptop with an Intel i7-10710U and 16GB of RAM. Bathtub models were coded in Python and run on a MacBook with an Apple M2 Pro chip and 16 GB of RAM. Dynamic traffic assignment for the two Delft networks took around 400s. The upper limit for the dynamic traffic assignment in each time step is five iterations. The M model simulation took about 0.2s, the event-based simulation took about 20s, and the trip-based simulations with fixed time steps took about 100s. 
\subsection{Comparison of models}
\begin{figure}[h!]
    \centering
    \begin{subfigure}[b]{0.54\textwidth}
        \centering
        \includegraphics[width=\textwidth]{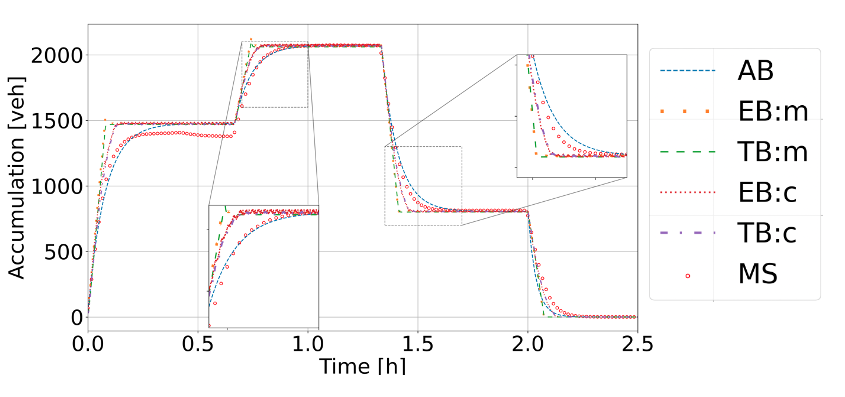}
        \caption{DF-D-1}
        \label{fig:del_dy4}
    \end{subfigure}
    \hfill
    \begin{subfigure}[b]{0.44\textwidth}
        \centering
        \includegraphics[width=\textwidth]{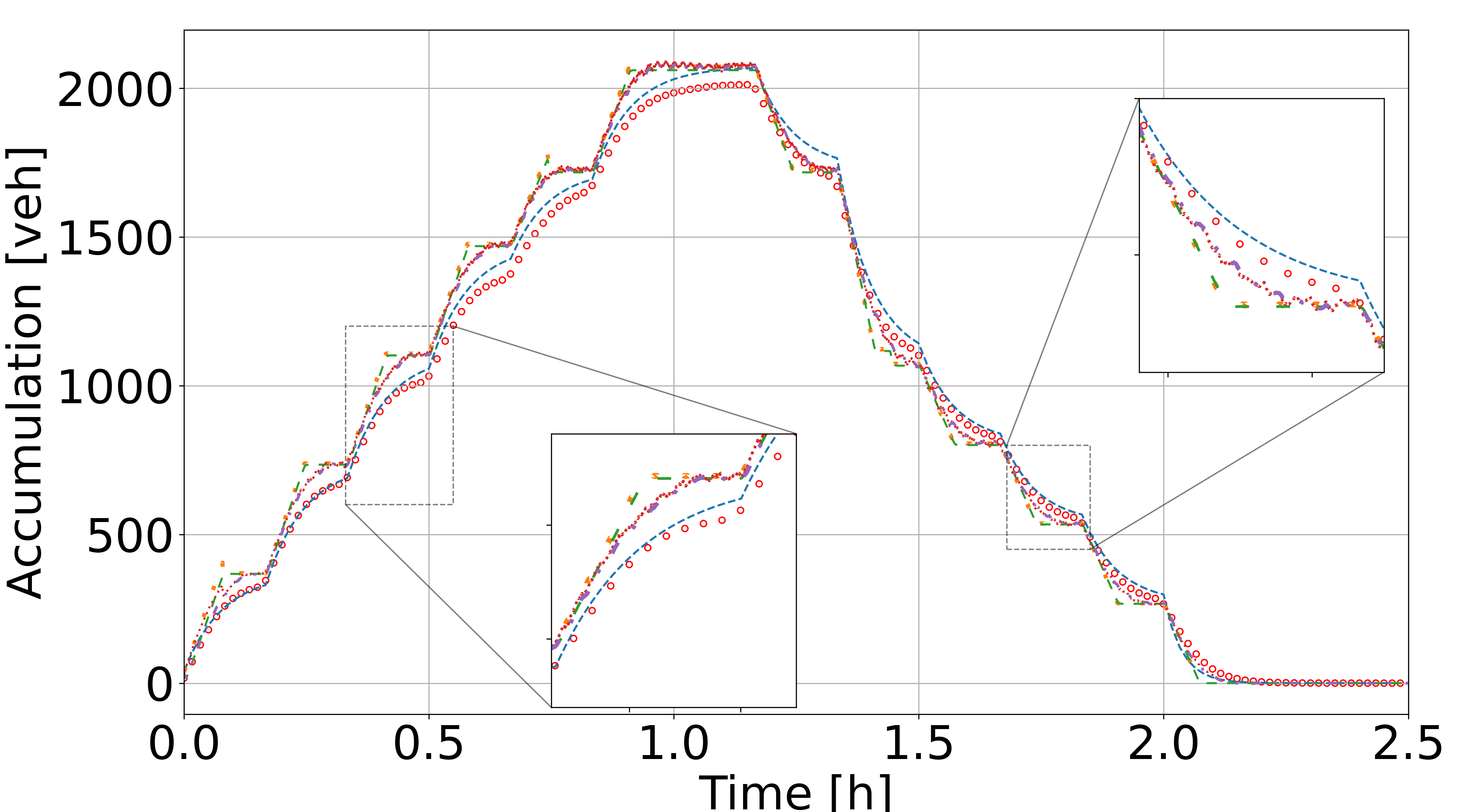}
        \caption{DF-D-2}
        \label{fig:del_dy12}
    \end{subfigure}
    \vspace{0.5cm}
    \begin{subfigure}[b]{0.48\textwidth}
        \centering
        \includegraphics[width=\textwidth]{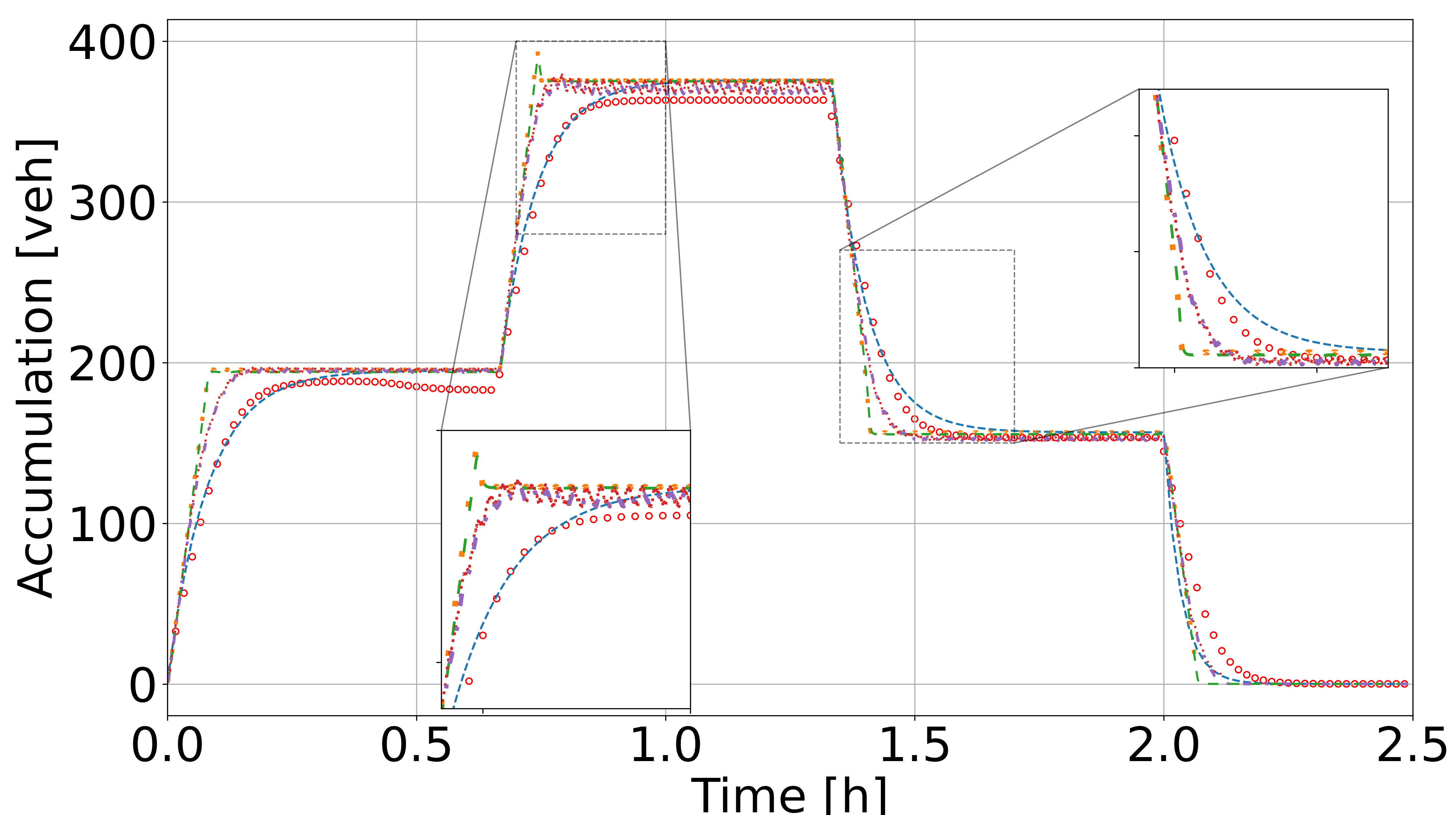}
        \caption{DU-D-1}
        \label{fig:delint_dyn_4}
    \end{subfigure}
    \hfill
    \begin{subfigure}[b]{0.48\textwidth}
        \centering
        \includegraphics[width=\textwidth]{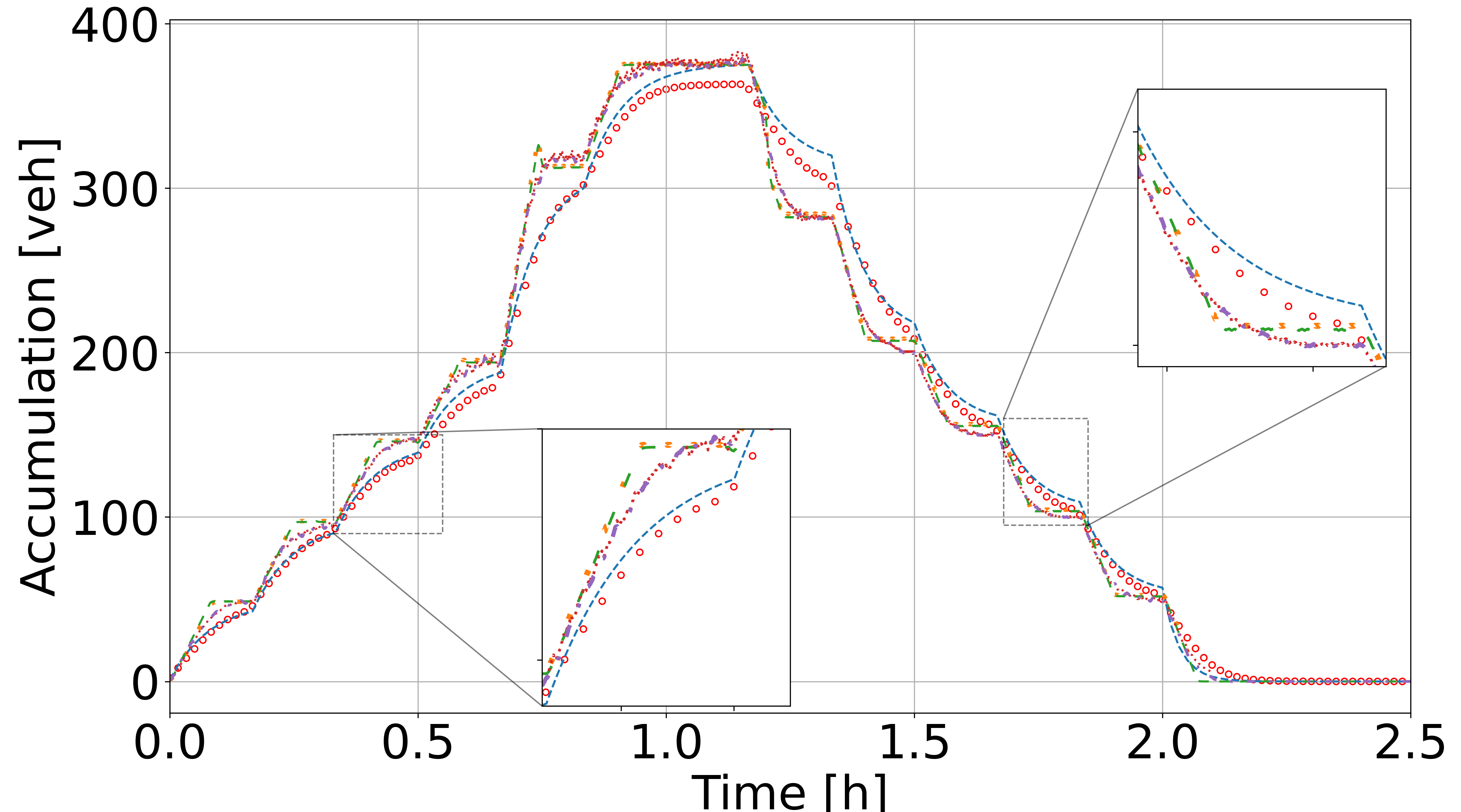}
        \caption{DU-D-2}
        \label{fig:delint_dyn_12}
    \end{subfigure}

    \caption{Accumulation comparison between models in different scenarios with dynamic TDD. (a,b) Delft network with freeways (DF); (c,d) Delft urban network (DU).}
    \label{fig:main_dyn}
\end{figure}

% In this section, results from different models will be referred to in abbreviations for convenience. In the following figures in this section. The accumulation-based model is referred to as "AB". Event-based simulations are referred to as "EB:m"(mean trip distance) and "EB:c"(trip distance categories). Similarly, AB2M simulations are referred to as "AB2M:m" and "AB2M:c". "Macrosimulation" will be used to represent results from the macroscopic traffic simulation in OmniTrans. 
Figures \ref{fig:main} and \ref{fig:main_dyn} illustrate the accumulation data from macroscopic traffic simulation and different bathtub models. Each sub-figure contains six sets of data representing the macroscopic traffic simulation data (MS), results by event-based simulation (EB) and trip-based model with fixed time steps (TB) using average trip distance (EB:m and TB:m) and using trip-distance categories (EB:c and TB:c), and results by the M model (AB). The results of the accumulation-based simulation are not in the figures because the optimal $\alpha$ for every scenario is 0. This means the M model simulation is equivalent to the accumulation-based model. It is referred to as "AB" in the following discussion.
%Some demand-changing parts of the curves are zoomed in to show the detailed differences between curves.

\begin{figure}[h!]
    \centering
    \begin{subfigure}[b]{0.54\textwidth}
        \centering
        \includegraphics[width=\textwidth]{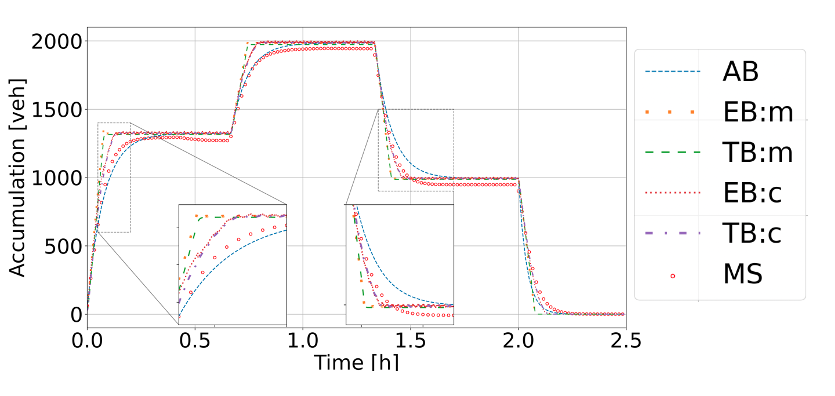}
        \caption{DF-S-1}
        \label{fig:del_st4}
    \end{subfigure}
    \hfill
    \begin{subfigure}[b]{0.44\textwidth}
        \centering
        \includegraphics[width=\textwidth]{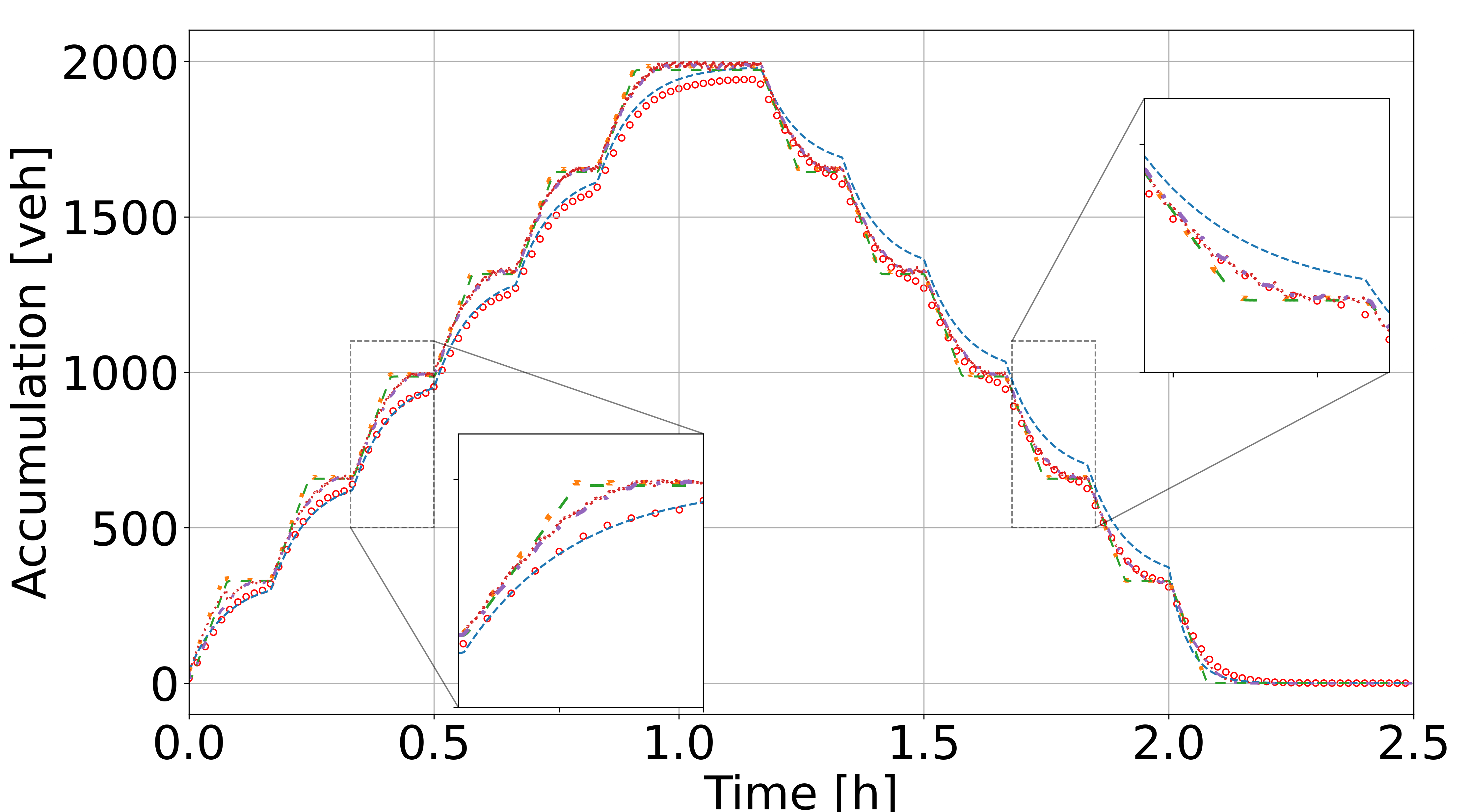}
        \caption{DF-S-2}
        \label{fig:del_st_12}
    \end{subfigure}
    \vspace{0.5cm}
    \begin{subfigure}[b]{0.48\textwidth}
        \centering
        \includegraphics[width=\textwidth]{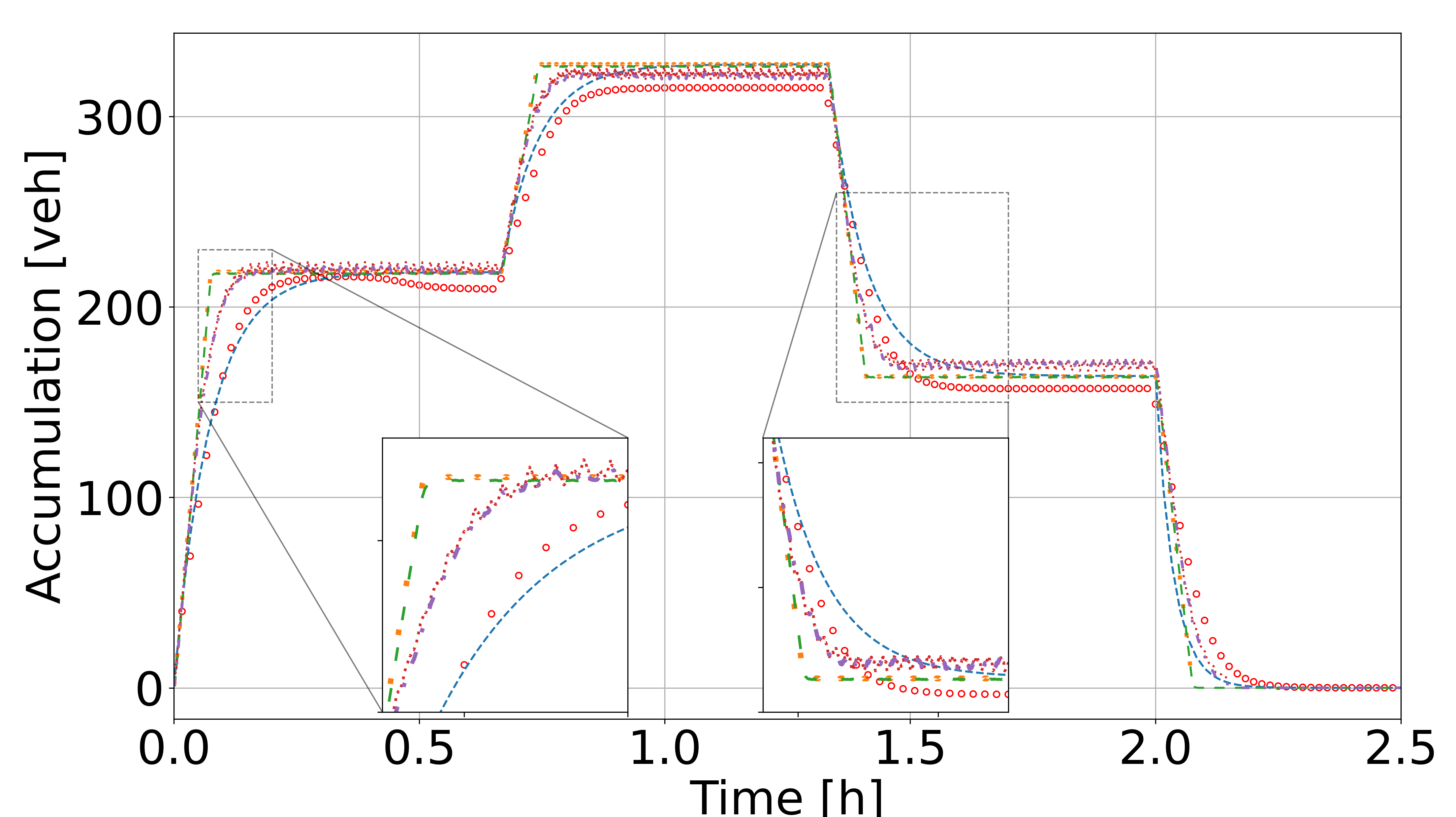}
        \caption{DU-S-1}
        \label{fig:delint_sta_4}
    \end{subfigure}
    \hfill
    \begin{subfigure}[b]{0.48\textwidth}
        \centering
        \includegraphics[width=\textwidth]{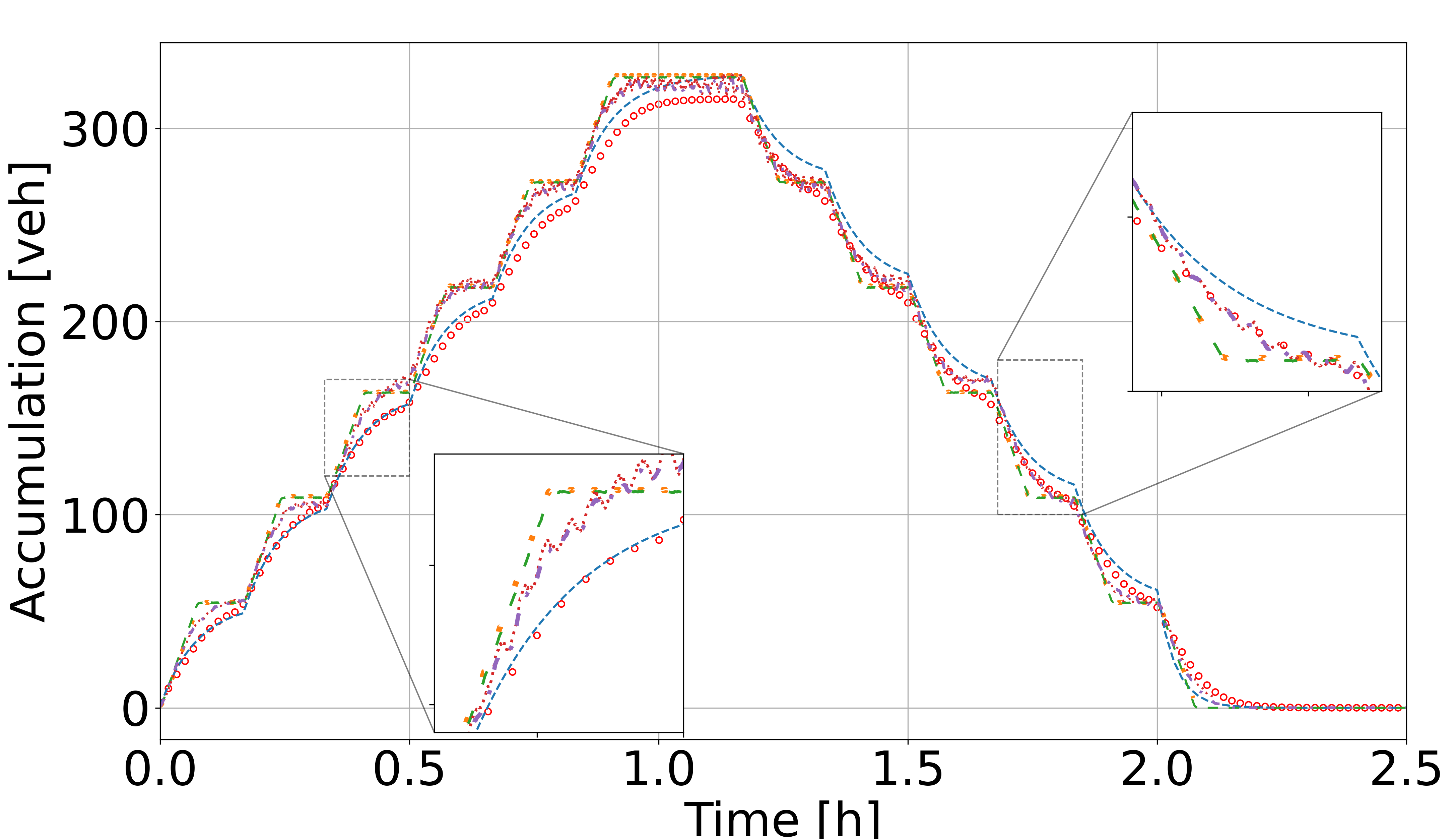}
        \caption{DU-S-2}
        \label{fig:delint_sta_12}
    \end{subfigure}
    \vspace{0.5cm}
    \begin{subfigure}[b]{0.48\textwidth}
        \centering
        \includegraphics[width=\textwidth]{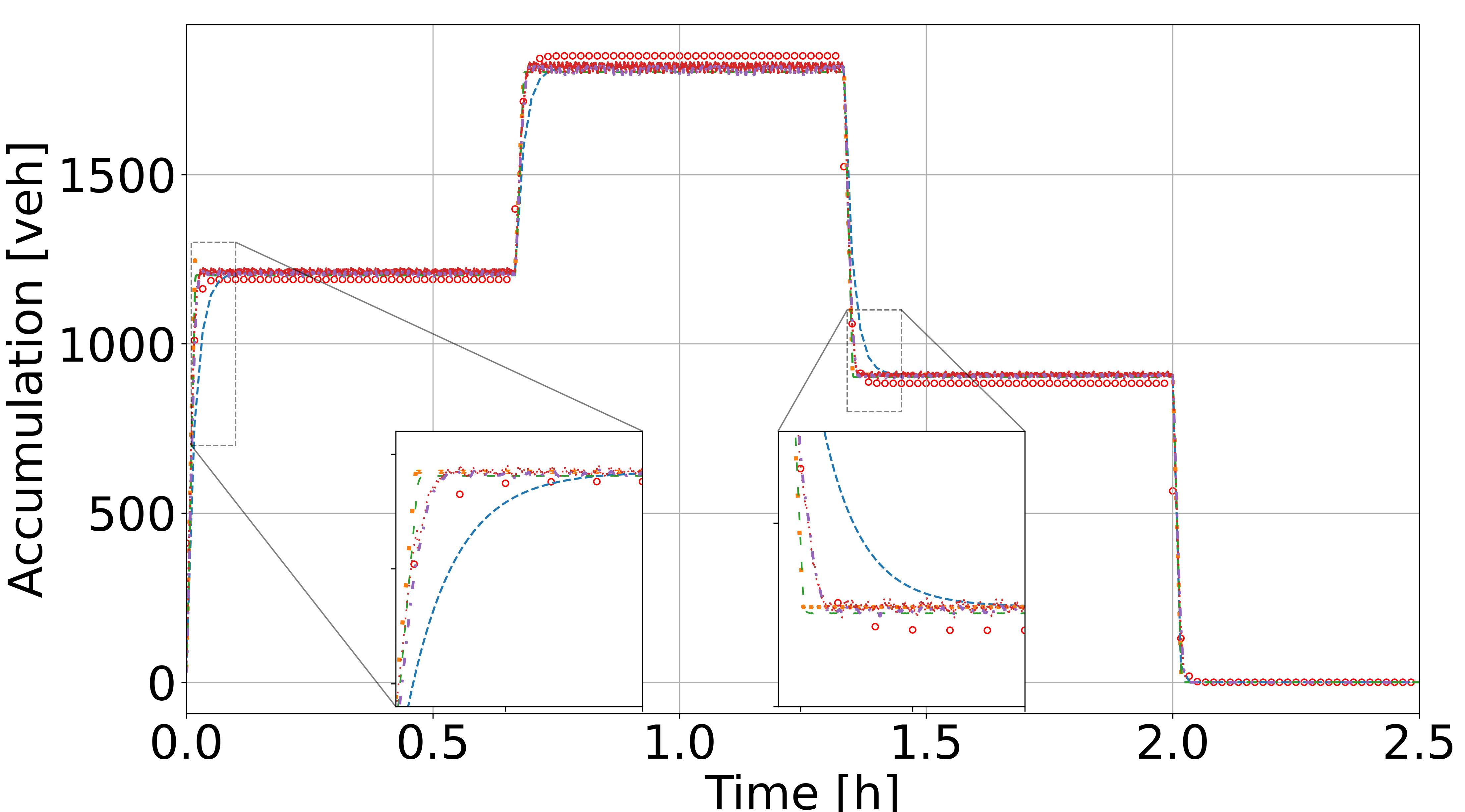}
        \caption{T-S-1}
        \label{fig:simp_st_4}
    \end{subfigure}
    \hfill
    \begin{subfigure}[b]{0.48\textwidth}
        \centering
        \includegraphics[width=\textwidth]{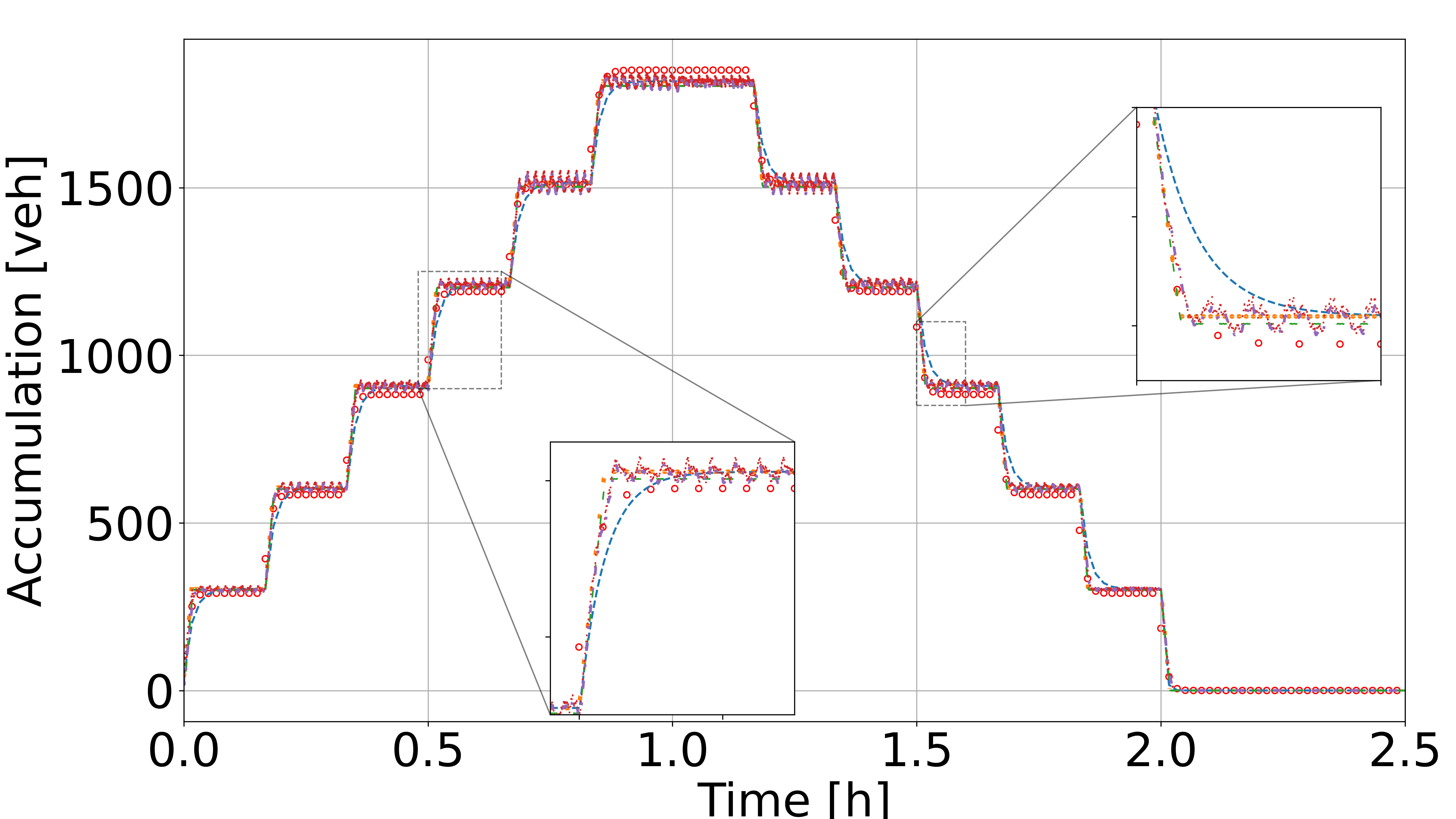}
        \caption{T-S-2}
        \label{fig:simp_st_12}
    \end{subfigure}
    
    \caption{Accumulation comparison between models in different scenarios with static TDD. (a,b) Delft network with freeways (DF); (c,d) Delft urban network (DU); (e,f) Toy network (T).}
    \label{fig:main}
\end{figure}

Using TDD at different aggregation levels has a significant effect during the transition states. With the TDD in the DF and DU networks, two trip-based simulations both have less steep increases and decreases when using trip distance categories than when using the MTD. This can be shown more obviously in scenarios with fast varying demand profiles (Figures \ref{fig:del_st4}, \ref{fig:delint_sta_4}, \ref{fig:del_dy4}, and \ref{fig:delint_dyn_4}). In the toy network, the difference between using trip distance categories and using the mean is relatively small for trip-based simulations, as shown in Figures \ref{fig:simp_st_4} and \ref{fig:simp_st_12}. This can be caused by the low trip distance variance. All the trip-based curves usually have steeper increases and decreases than the accumulation-based model. This has also been observed in other comparison studies \cite{Mariotte2017, Leclercq2019, huang-2024}. The reason for this is that the newly added trips in demand change can increase or decrease the speed (hence the outflow) of trips inside the reservoir in the trip-based models. In the accumulation-based model, the remaining trip distances are not represented, leading to higher outflow during demand surges and lower outflow during demand decreases. 

When looking at the steady states, different aggregation levels of TDD do not considerably affect the steady accumulation, as in all cases, the accumulations in steady states of the four trip-based models overlap with the curve of the accumulation-based model. The difference between bathtub models and macroscopic simulation is caused by the error in the MFD estimation, which results from speed heterogeneity in the networks. The slight difference between the event-based simulation and trip-based simulation with fixed time steps is because of the different time steps they used for updating the average speed in the network. 

Compared with the macroscopic traffic simulation data, it can be seen in the static TDD scenarios in both Delft networks (Figures \ref{fig:del_st4}, \ref{fig:del_st_12}, \ref{fig:delint_sta_4}, \ref{fig:delint_sta_12}) that the accumulation-based model is closer to the data during the demand increase and the trip-based simulations are closer during the demand decrease. In the dynamic TDD scenarios  (Figures \ref{fig:del_dy4}, \ref{fig:del_dy12}, \ref{fig:delint_dyn_4}, \ref{fig:delint_dyn_12}), the accumulation-based model fits better with both demand increase and decrease. However, in the toy network, trip-based simulations fit the data better. 

\begin{figure}[h!]
    \centering
    \begin{subfigure}[b]{0.48\textwidth}
        \centering
        \includegraphics[width=\textwidth]{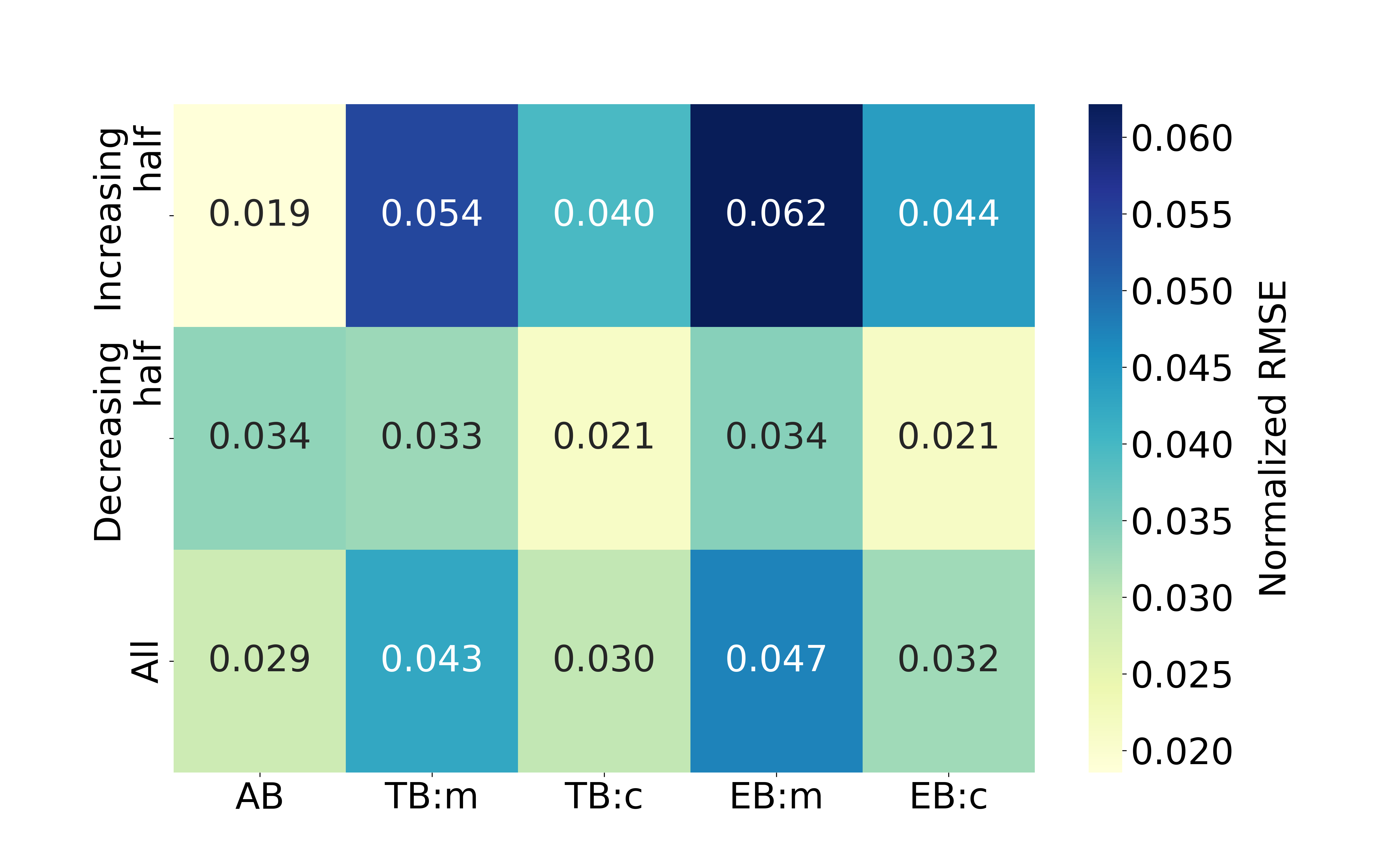}
        \caption{DF-S-1}
        \label{fig:heatdel_st4}
    \end{subfigure}
    \hfill
    \begin{subfigure}[b]{0.48\textwidth}
        \centering
        \includegraphics[width=\textwidth]{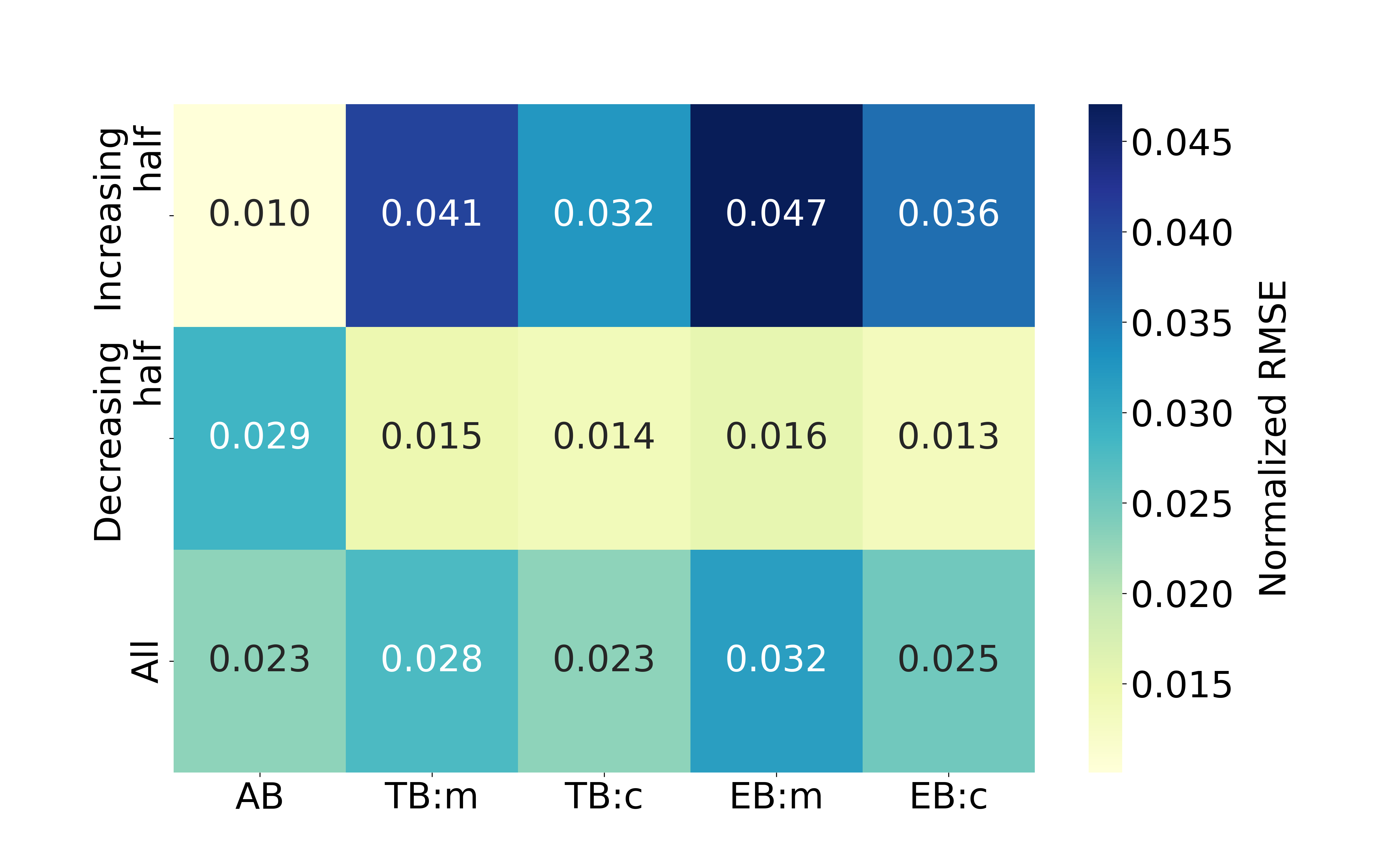}
        \caption{DF-S-2}
        \label{fig:heatdel_st_12}
    \end{subfigure}
    \vspace{0.5cm}
    \begin{subfigure}[b]{0.48\textwidth}
        \centering
        \includegraphics[width=\textwidth]{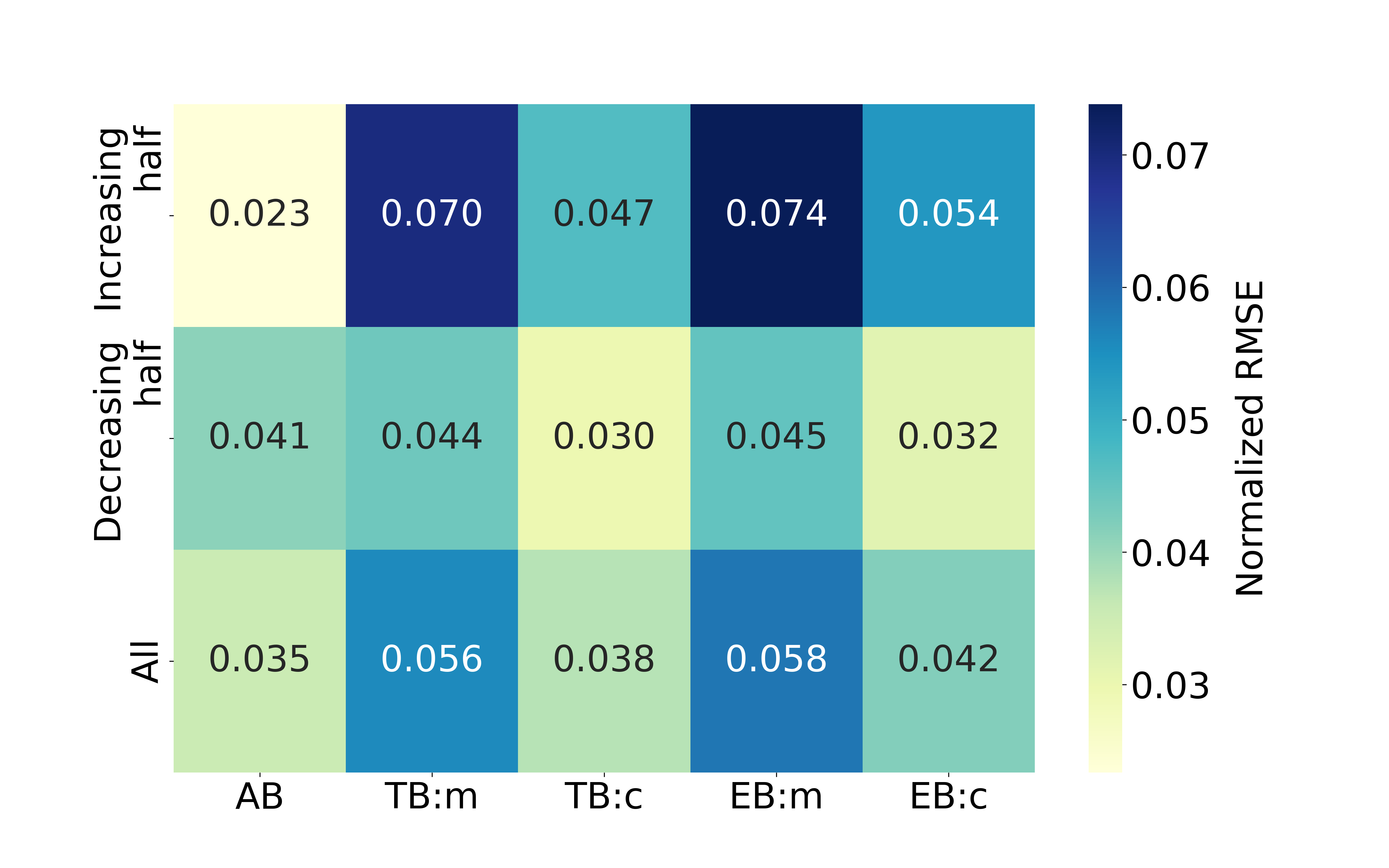}
        \caption{DU-S-1}
        \label{fig:heatdelint_st_4}
    \end{subfigure}
    \hfill
    \begin{subfigure}[b]{0.48\textwidth}
        \centering
        \includegraphics[width=\textwidth]{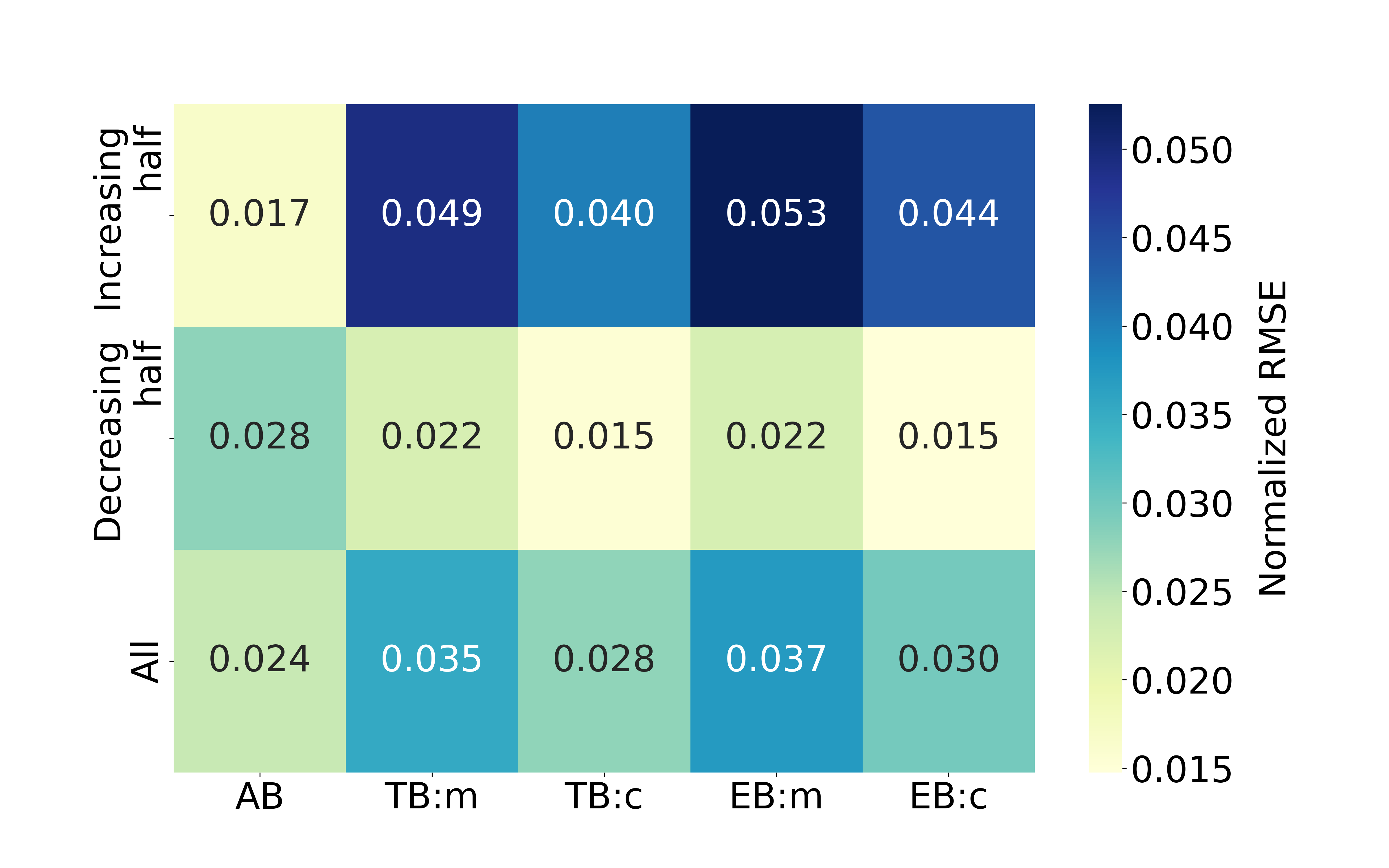}
        \caption{DU-S-2}
        \label{fig:heatdelint_st_12}
    \end{subfigure}
    \vspace{0.5cm}
    \begin{subfigure}[b]{0.48\textwidth}
        \centering
        \includegraphics[width=\textwidth]{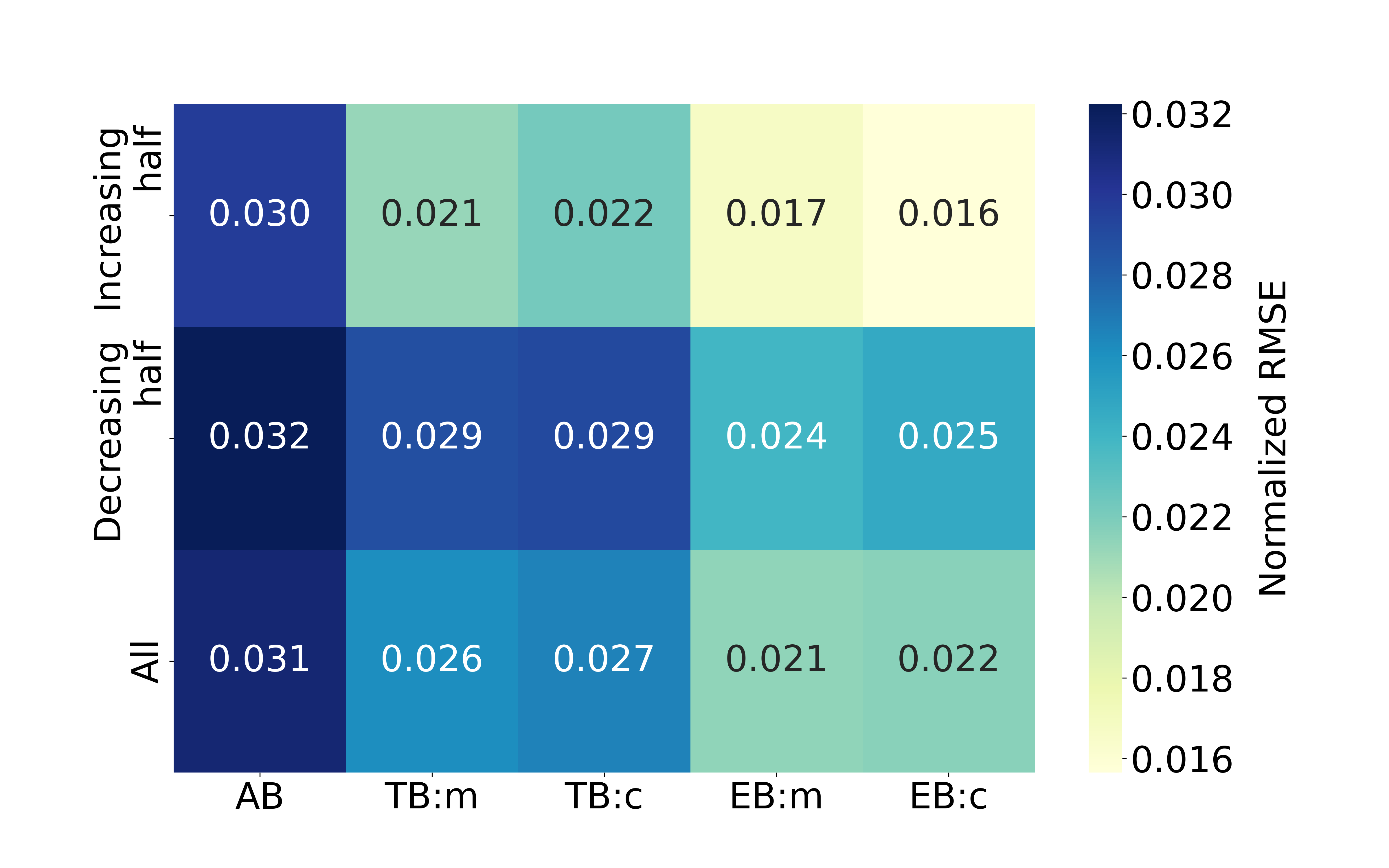}
        \caption{T-S-1}
        \label{fig:heatsimp_st_4}
    \end{subfigure}
    \hfill
    \begin{subfigure}[b]{0.48\textwidth}
        \centering
        \includegraphics[width=\textwidth]{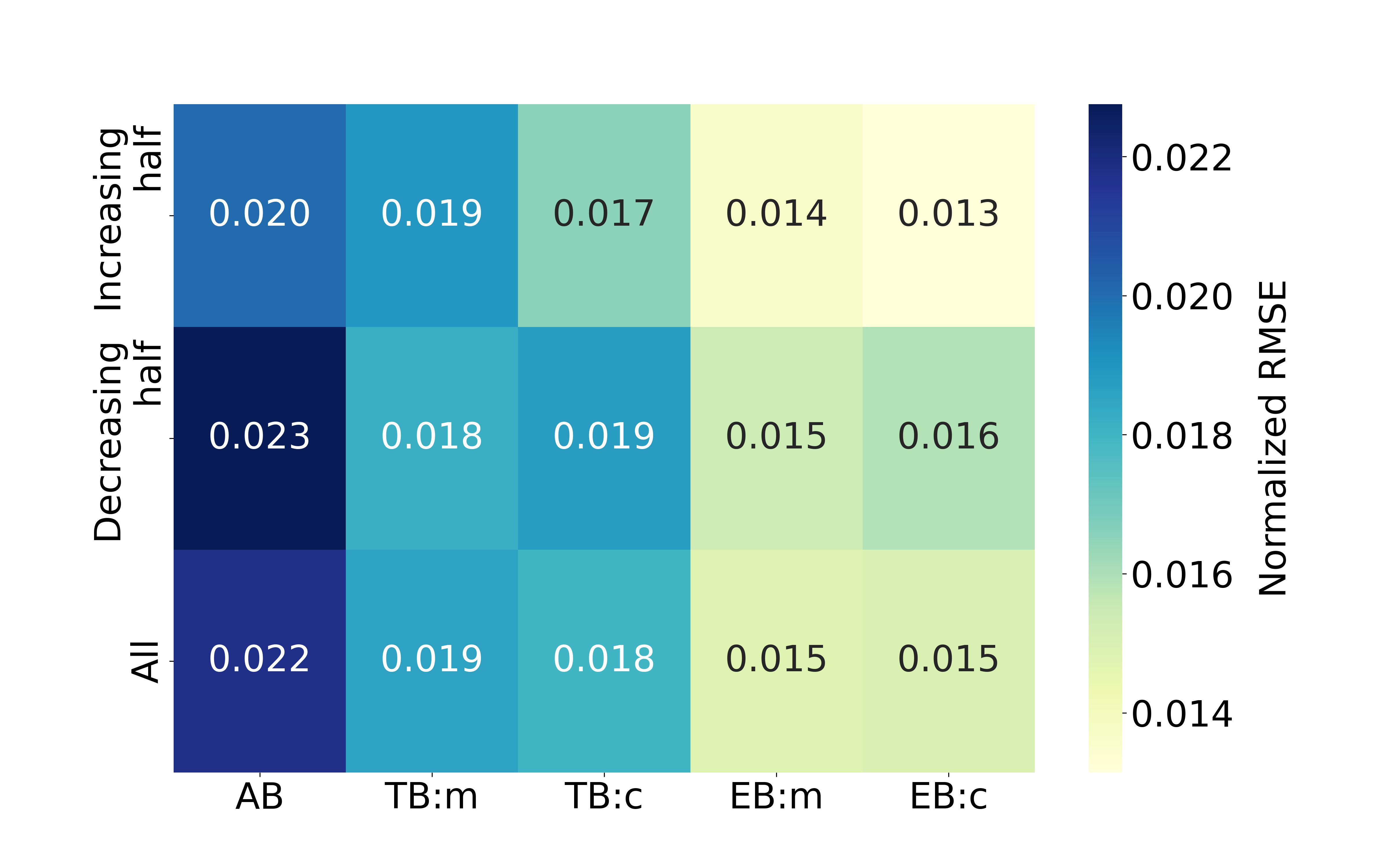}
        \caption{T-S-2}
        \label{fig:heatsimp_st_12}
    \end{subfigure}
    
    \caption{Normalised RMSE compared to macroscopic traffic simulation in different scenarios with static TDD. (a,b) Delft network with freeways (DF); (c,d) Delft urban network (DU); (e,f) Toy network (T).}
    \label{fig:heatmaps}
\end{figure}

\begin{figure}[h!]
    \centering
    \begin{subfigure}[b]{0.48\textwidth}
        \centering
        \includegraphics[width=\textwidth]{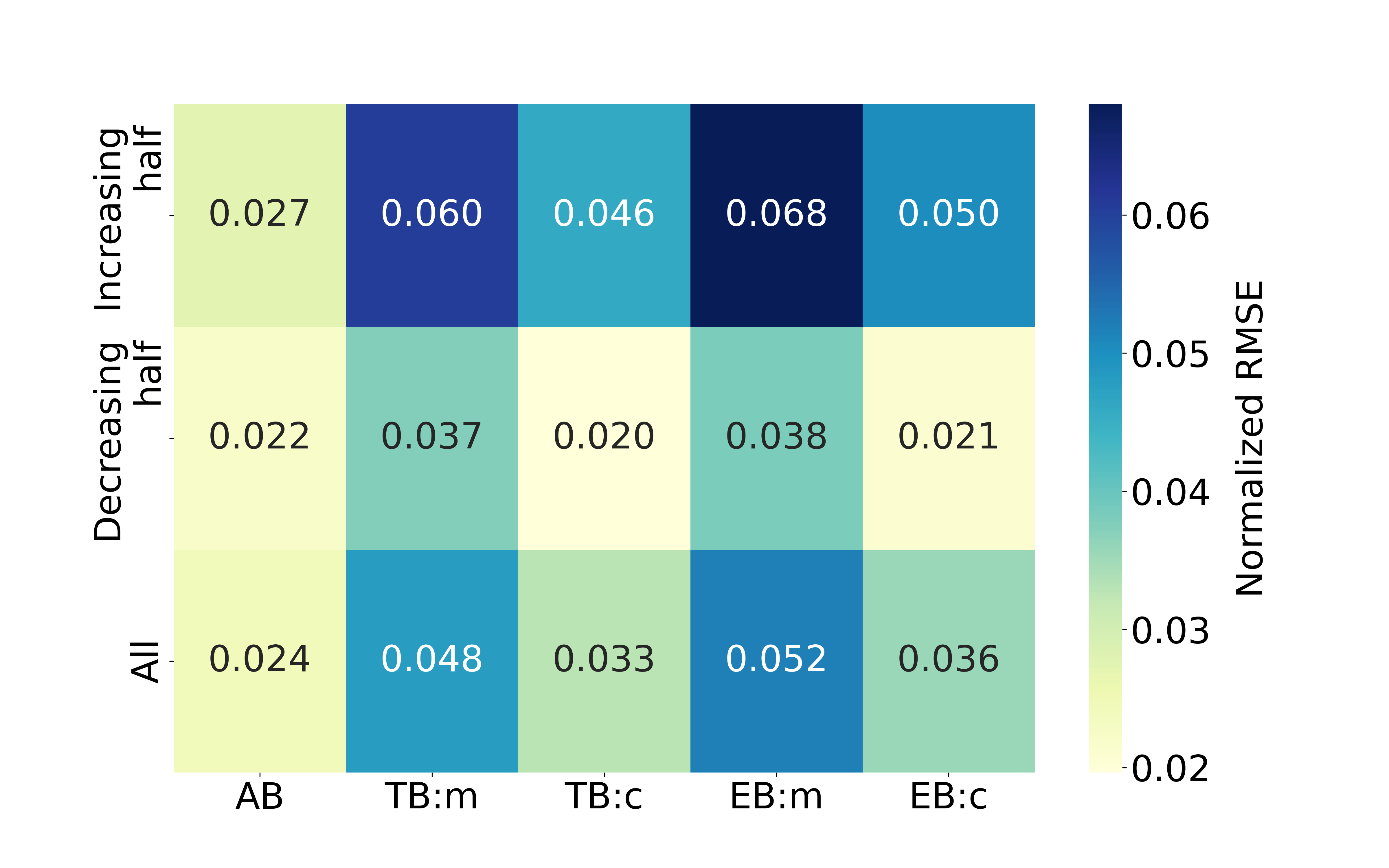}
        \caption{DF-D-1}
        \label{fig:heatdel_dy_4}
    \end{subfigure}
    \hfill
    \begin{subfigure}[b]{0.48\textwidth}
        \centering
        \includegraphics[width=\textwidth]{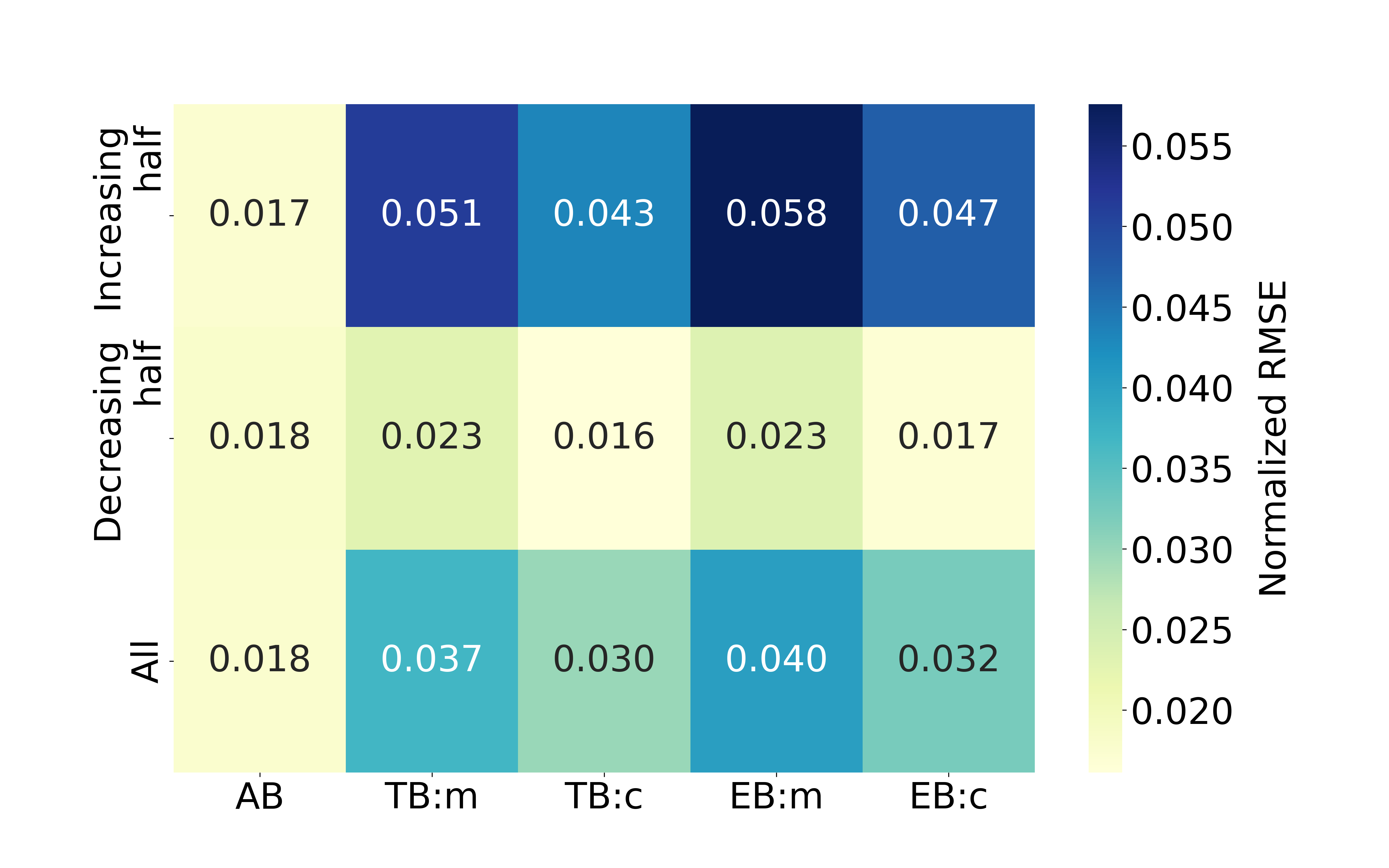}
        \caption{DF-D-2}
        \label{fig:heatdel_dy_12}
    \end{subfigure}
    \vspace{0.5cm}
    \begin{subfigure}[b]{0.48\textwidth}
        \centering
        \includegraphics[width=\textwidth]{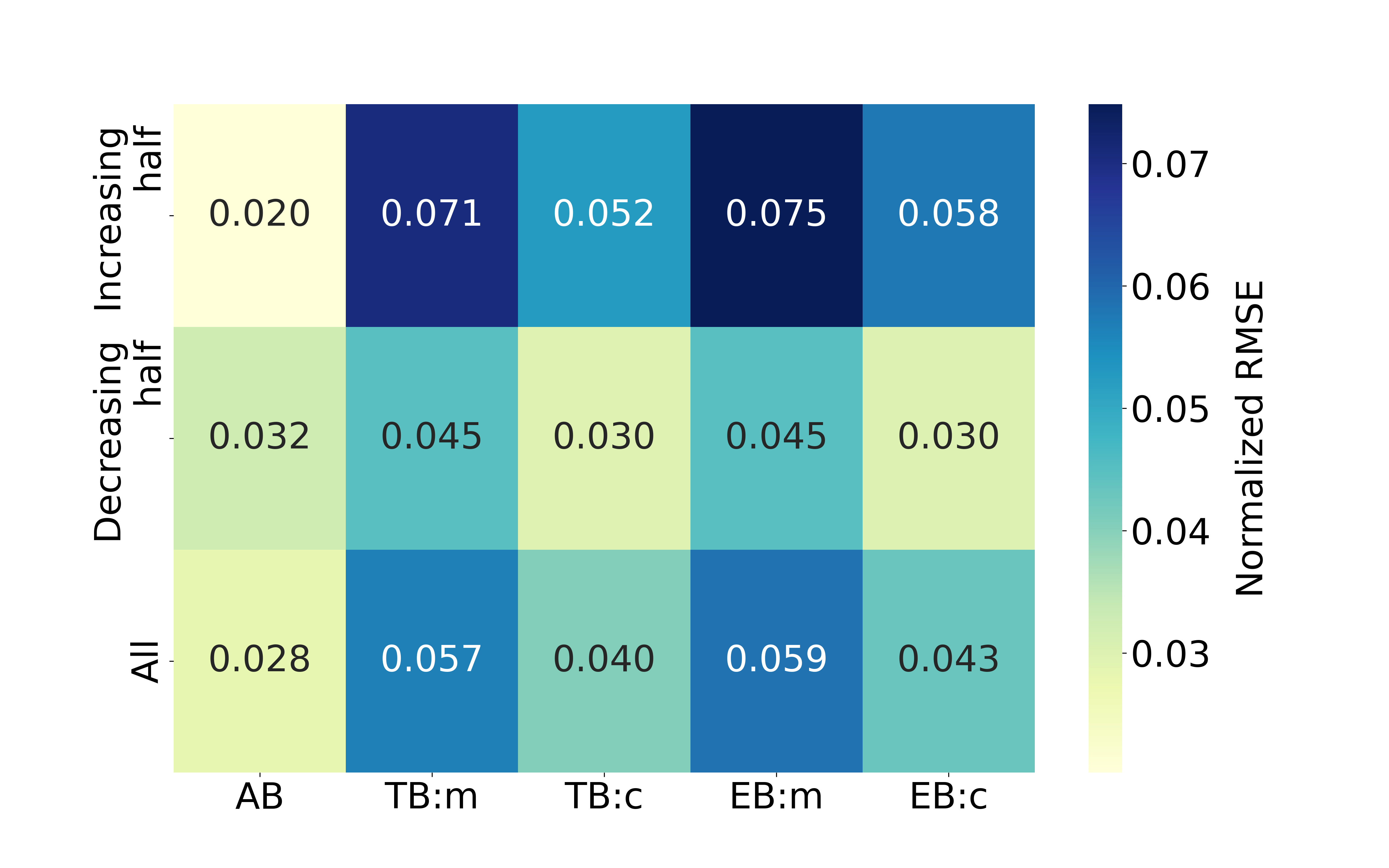}
        \caption{DU-D-1}
        \label{fig:heatdelint_dy_4}
    \end{subfigure}
    \hfill
    \begin{subfigure}[b]{0.48\textwidth}
        \centering
        \includegraphics[width=\textwidth]{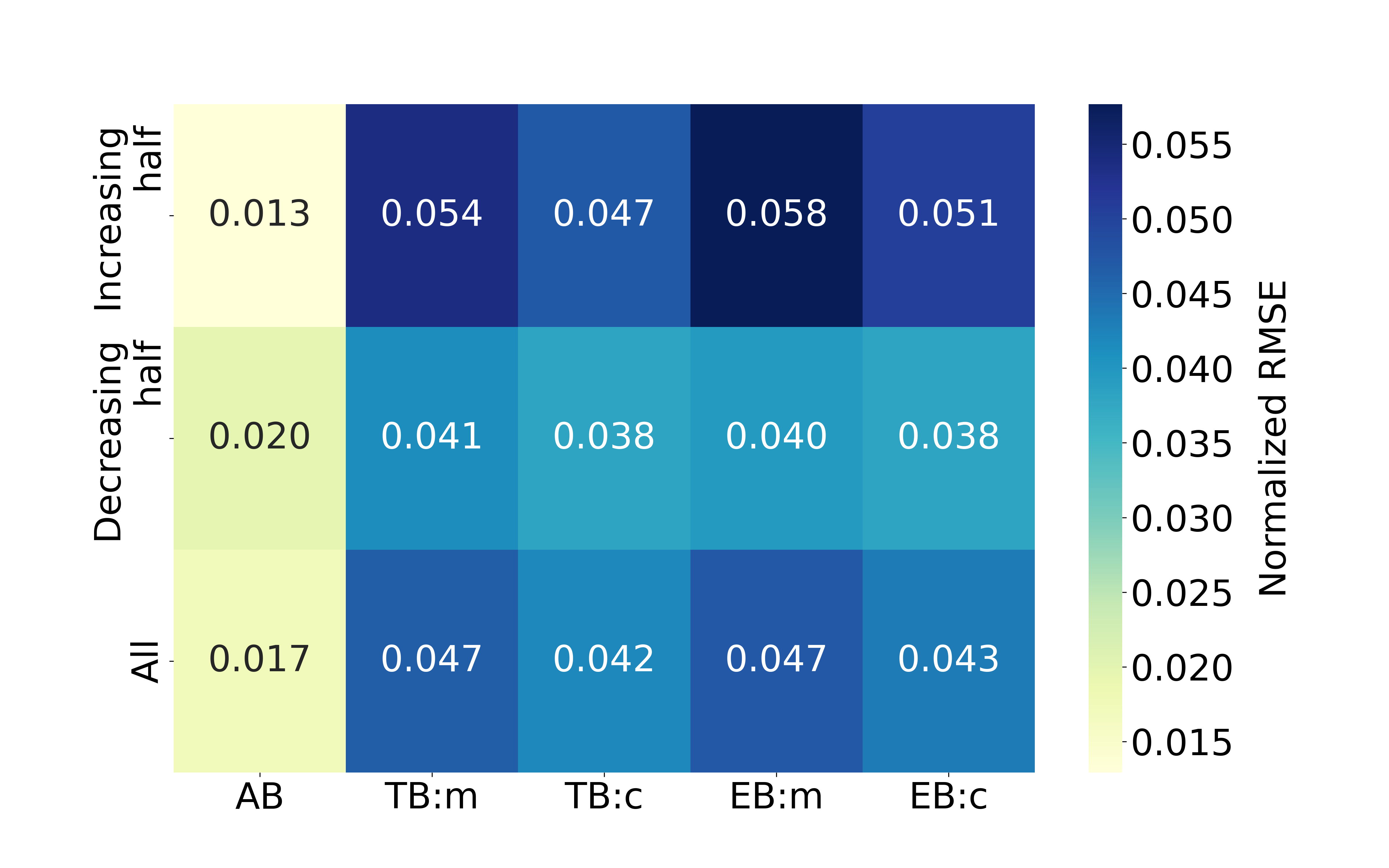}
        \caption{DU-D-2}
        \label{fig:heatdelint_dy_12}
    \end{subfigure}
    
    \caption{Normalised RMSE compared to macroscopic traffic simulation in different scenarios with dynamic TDD. (a,b) Delft network with freeways (DF); (c,d) Delft urban network (DU).}
    \label{fig:heatmaps_dyn}
\end{figure}

Numerical errors are calculated to have a more thorough understanding of the difference between macroscopic traffic simulation and bathtub models in different states. Normalized RMSE is used as the main indicator. The RMSE between the results of bathtub models and the macroscopic traffic simulation are normalized by the maximum accumulation. As the trip-based simulations have fluctuating results, data smoothing and aggregation are required to make them comparable with the macroscopic traffic simulation results. For all trip-based simulations, exponential smoothing is applied with a smoothing parameter of 0.2. 

Results of normalized RMSE are shown in heat maps in Figures \ref{fig:heatmaps} and \ref{fig:heatmaps_dyn}. 
The y-axis represents different data considered in RMSE calculation. Data from demand increasing half of the simulation (from 0 to 1h), the demand decreasing half (from 1h to 2.5h), and the whole simulation duration are compared. When comparing fast varying demand scenarios ( Figures \ref{fig:heatdel_st4}, \ref{fig:heatdelint_st_4}, \ref{fig:heatsimp_st_4}, \ref{fig:heatdel_dy_4}, and \ref{fig:heatdelint_dy_4}) with slow varying demand scenarios (Figures \ref{fig:heatdel_st_12}, \ref{fig:heatdelint_st_12}, \ref{fig:heatsimp_st_12}, \ref{fig:heatdel_dy_12}, and \ref{fig:heatdelint_dy_12}) it can be seen that the errors are in general smaller with the slow-varying demand (profile 2) than with demand profile 1. This shows that fast demand changes can lead to more errors in bathtub models. 

The effect of including dynamic TDD can be analyzed by comparing static TDD scenarios in two Delft networks (Figures \ref{fig:heatdel_st4}, \ref{fig:heatdel_st_12}, \ref{fig:heatdelint_st_4}, and \ref{fig:heatdelint_st_12}) with dynamic TDD scenarios (Figures \ref{fig:heatdel_dy_4}, \ref{fig:heatdel_dy_12}, \ref{fig:heatdelint_dy_4}, and \ref{fig:heatdelint_dy_12}). The normalized RMSE increases in the dynamic TDD scenarios in all trip-based models while it decreases in the accumulation-based model. This means the accumulation-based model can adapt to dynamic TDDs better than trip-based models in the two Delft networks. 

With higher density homogeneity, scenarios in the DU network are expected to have lower errors than the DF network. However, this is not the case when comparing Figures \ref{fig:heatdel_st4}, \ref{fig:heatdel_st_12}, \ref{fig:heatdel_dy_4}, and \ref{fig:heatdel_dy_12} with Figures \ref{fig:heatdelint_st_4}, \ref{fig:heatdelint_st_12}, \ref{fig:heatdelint_dy_4}, and \ref{fig:heatdelint_dy_12}. The higher errors in the DU network may be caused by the error in MFD estimation. Lower trip numbers may also exaggerate the errors despite that the errors are normalized. 

\begin{figure}[h!]
    \centering
    \begin{subfigure}[b]{0.52\textwidth}
        \centering
        \includegraphics[width=\textwidth]{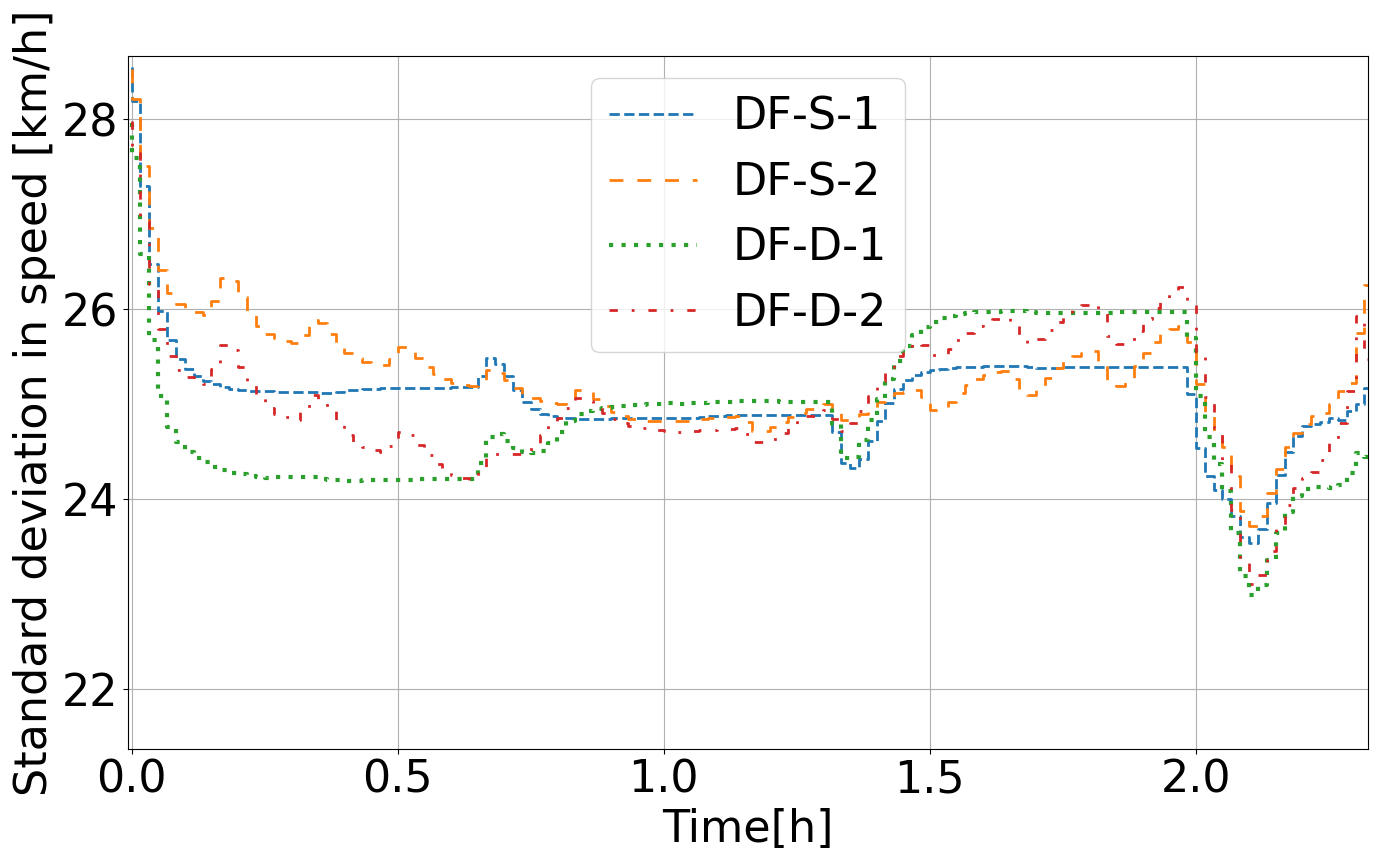}
        \caption{}
        \label{fig:spddev_del}
    \end{subfigure}
    \vspace{0.5cm}
    \begin{subfigure}[b]{0.52\textwidth}
        \centering
        \includegraphics[width=\textwidth]{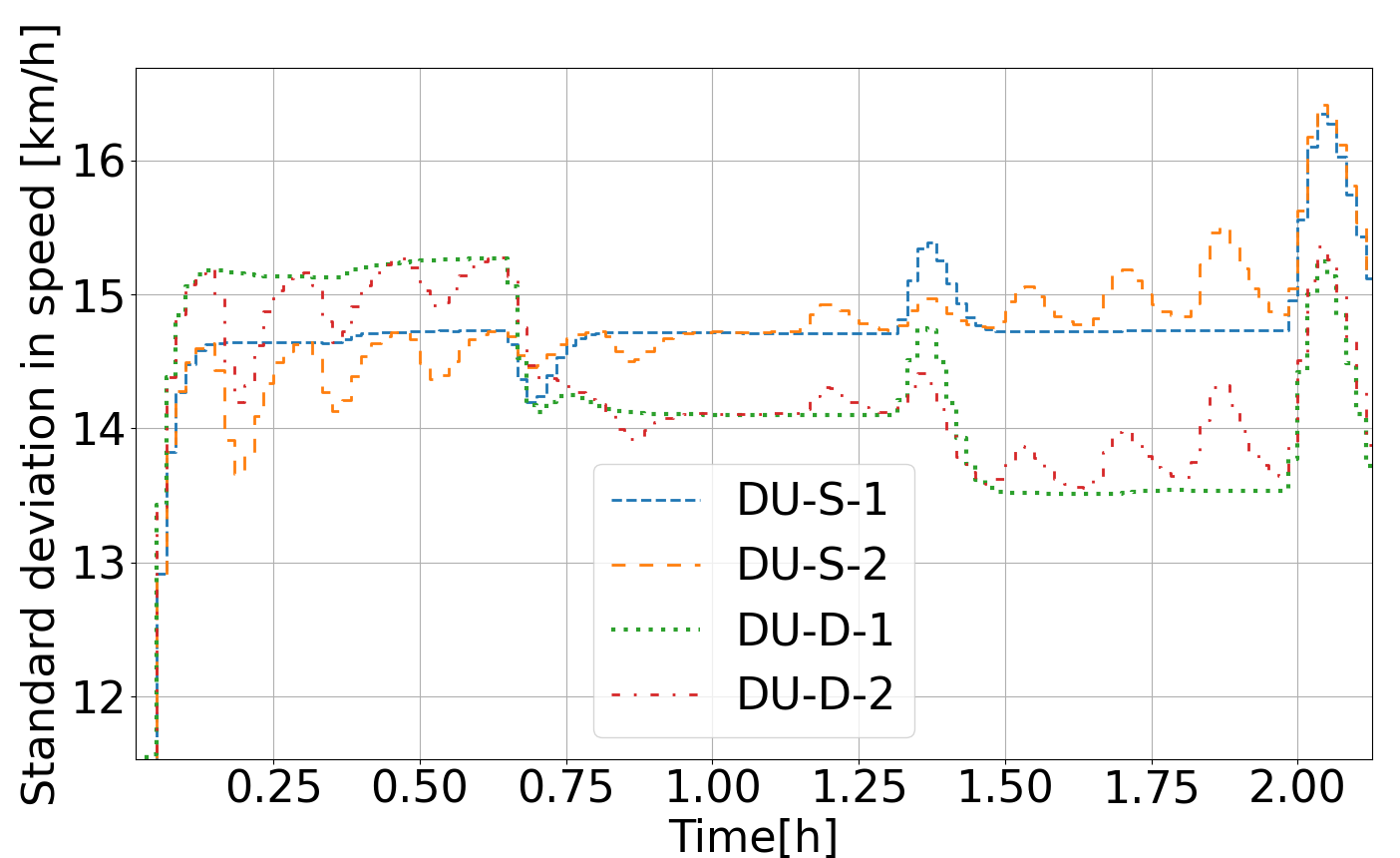}
        \caption{}
        \label{fig:spddev_delint}
    \end{subfigure}
    \vspace{0.5cm}
    \begin{subfigure}[b]{0.52\textwidth}
        \centering
        \includegraphics[width=\textwidth]{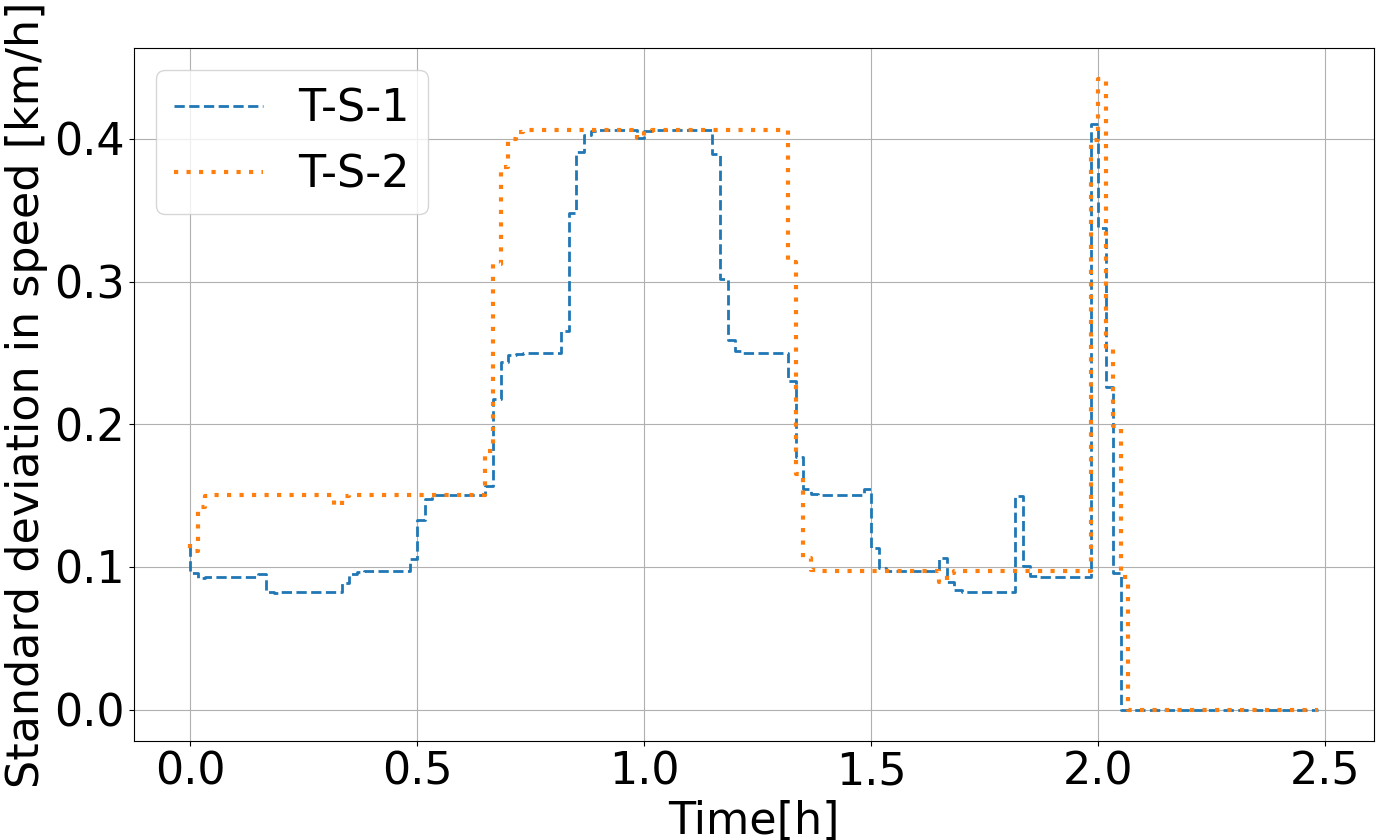}
        \caption{}
        \label{fig:spddev_simp}
    \end{subfigure}

    \caption{Temporal evolution of standard deviation in speed in three network cases. (a) Delft network with freeways (DF); (b) Delft urban network (DU); (c) Toy network (T).}
    \label{fig:speed dev}
\end{figure}

As the performance of bathtub models differs considerably among scenarios, the temporal evolution of the speed standard deviation during the simulation is explored to understand how the three networks respond to demand and TDD changes. As shown in Figure \ref{fig:speed dev}, the DF network has a speed standard deviation twice higher than the DU network due to including the freeways. The toy network has the highest speed homogeneity and only a speed standard deviation less than 0.5. It can be seen in Figures \ref{fig:spddev_del} and \ref{fig:spddev_delint} that it takes longer for both DF and DU networks to reach steady states. This delay in the state transitions compensates for the slow response of the accumulation-based model to demand changes, making it the best fit in all scenarios in the DF network. For scenarios in the toy network, the speed deviation can rapidly reach a new steady state. The difference in the state transition speed in the two Delft networks and the toy network case could be caused by the difference in the network sizes and the spatial demand distribution.

% The network loading time does not affect the model results equally during demand increases and decreases. \citep{mousavizadeh-2024} found that network properties are more influential for MFD estimation during the network loading phase than during the unloading phase. The differences in RMSE between demand increasing and decreasing half we found in simulations are in line with their findings.

% It takes time for the increased or decreased demand to fill the network and reach a new accumulation equilibrium. For the toy network, the demand change can rapidly affect the whole network. It takes much longer for the Delft network to reach a new equilibrium. 

% This property makes the accumulation change even more slowly since vehicles need to travel further to fill the network. 

% \subsection{Sensitivity analysis on time resolution for AB2M simulation}

% \begin{figure}[htbp]
%     \centering
%     \begin{subfigure}[b]{0.48\textwidth}
%         \centering
%         \includegraphics[width=\textwidth]{fig/Sensitivity_cat.png}
%         \caption{}
%         \label{fig:sensitivity cat}
%     \end{subfigure}
%     \hfill
%     \begin{subfigure}[b]{0.48\textwidth}
%         \centering
%         \includegraphics[width=\textwidth]{fig/Sensitivity_mean.png}
%         \caption{}
%         \label{fig:sensitivity mean}
%     \end{subfigure}

%     \caption{Observation on the relationship between numerical error and normalised RMSE of AB2M simulation. (a) AB2M with trip distance categories. The relationship between error in production and normalized RMSE; (b) AB2M with mean trip distances. The relationship between error in trip number and normalized RMSE.}
%     \label{fig:main}
% \end{figure}

\subsection{Hysteresis loops}
\begin{figure}[htbp]
    \centering
    \begin{subfigure}[t]{0.48\textwidth}
        \centering
        \includegraphics[width=\textwidth]{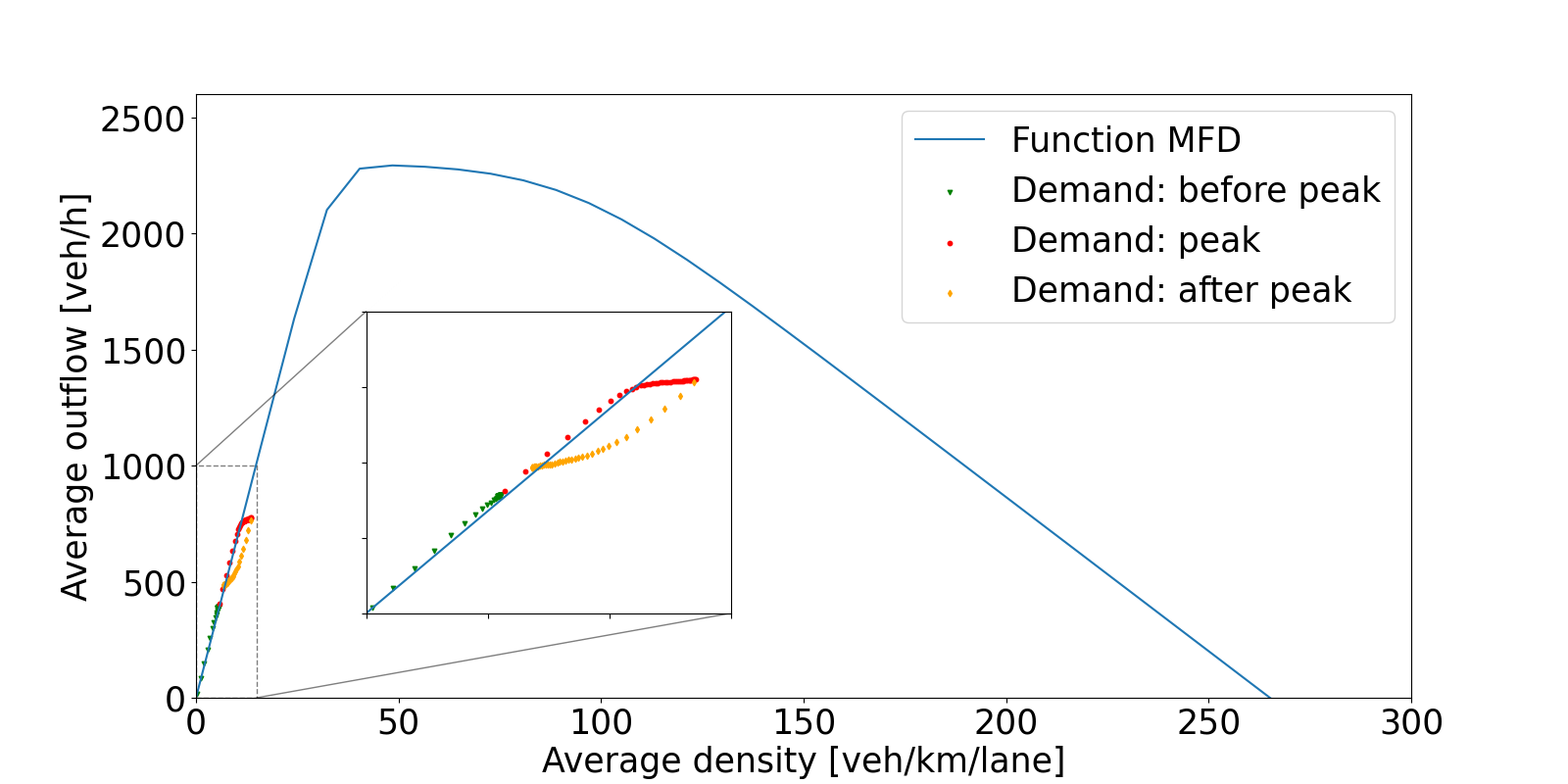}
        \caption{}
        \label{fig:hyst}
    \end{subfigure}
    \hfill
    \begin{subfigure}[t]{0.48\textwidth}
        \centering
        \includegraphics[width=\textwidth]{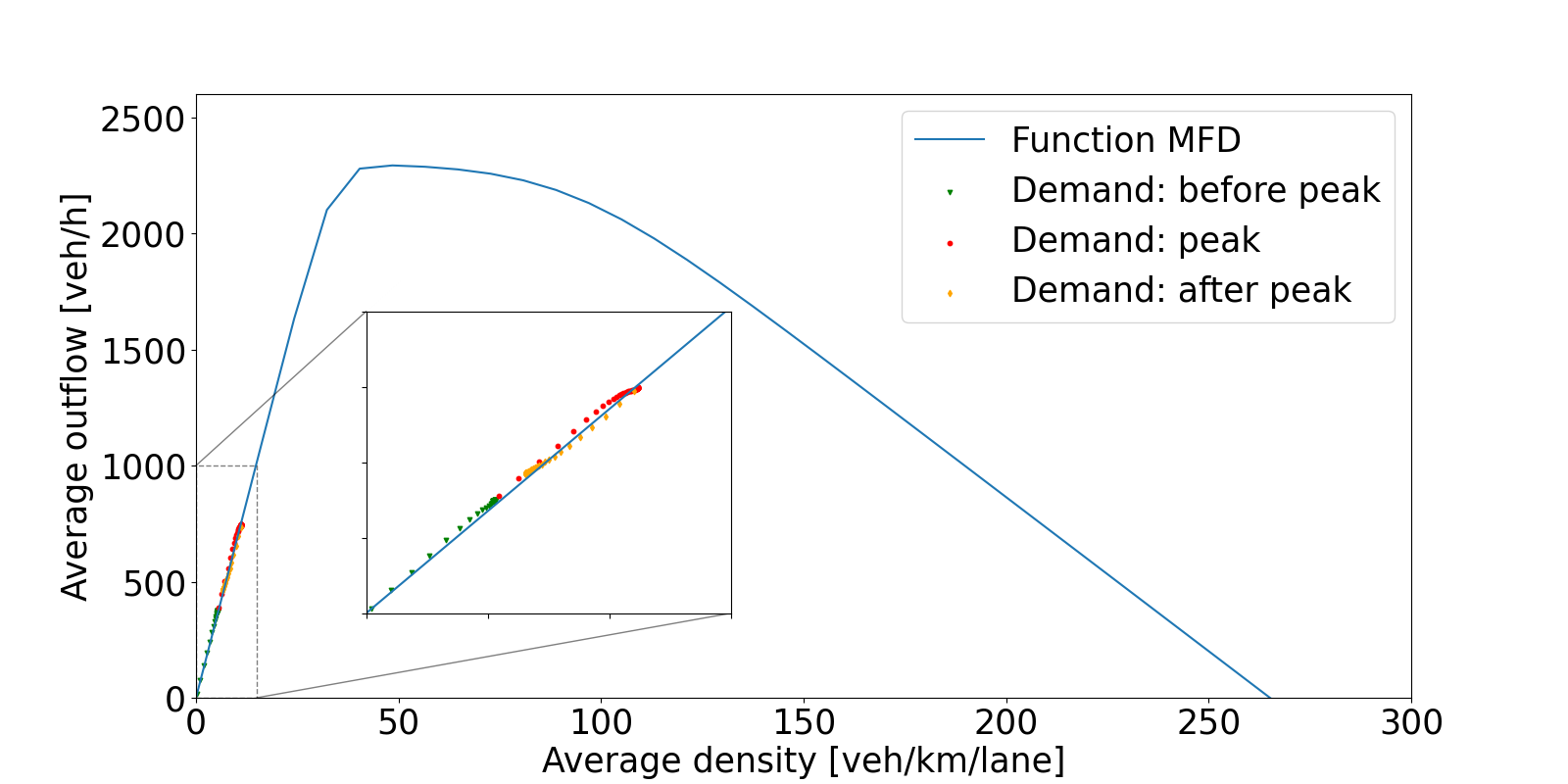}
        \caption{}
        \label{fig:remove bottleneck}
    \end{subfigure}
    \vspace{0.5cm}
    \begin{subfigure}[t]{0.46\textwidth}
        \centering
        \includegraphics[width=\textwidth]{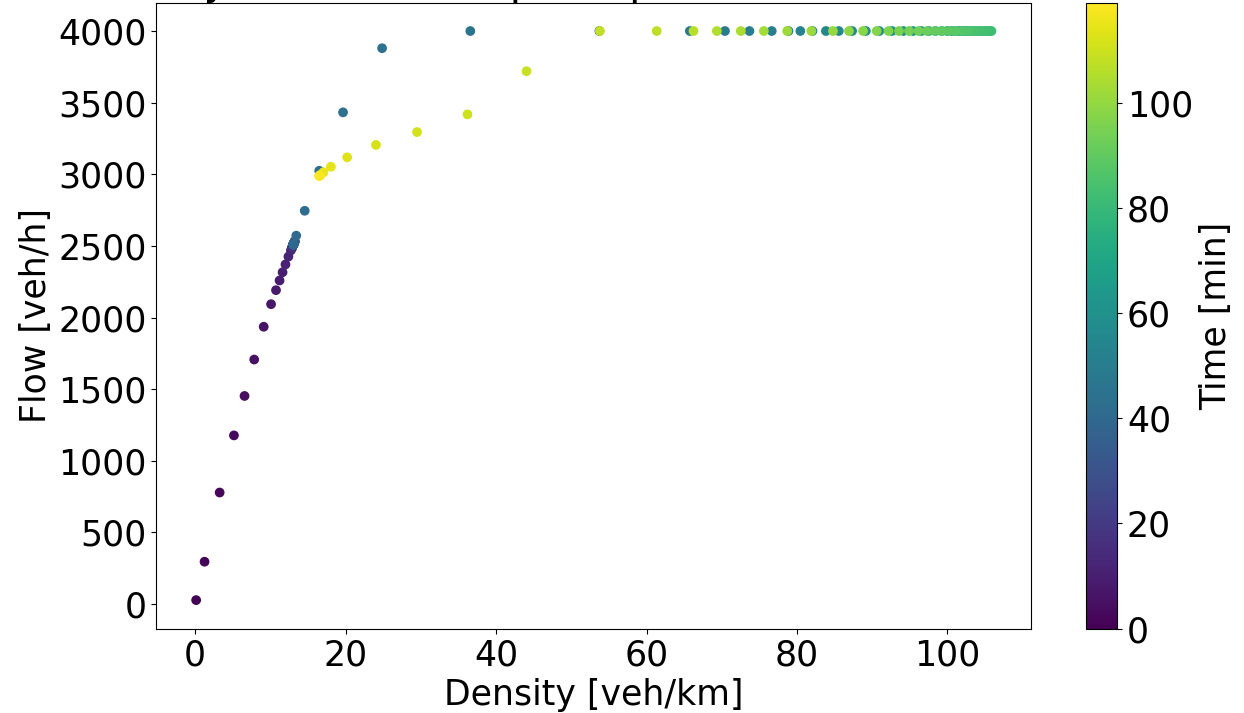}
        \caption{}
        \label{fig:link3lane}
    \end{subfigure}
    \hfill
    \begin{subfigure}[t]{0.48\textwidth}
        \centering
        \includegraphics[width=\textwidth]{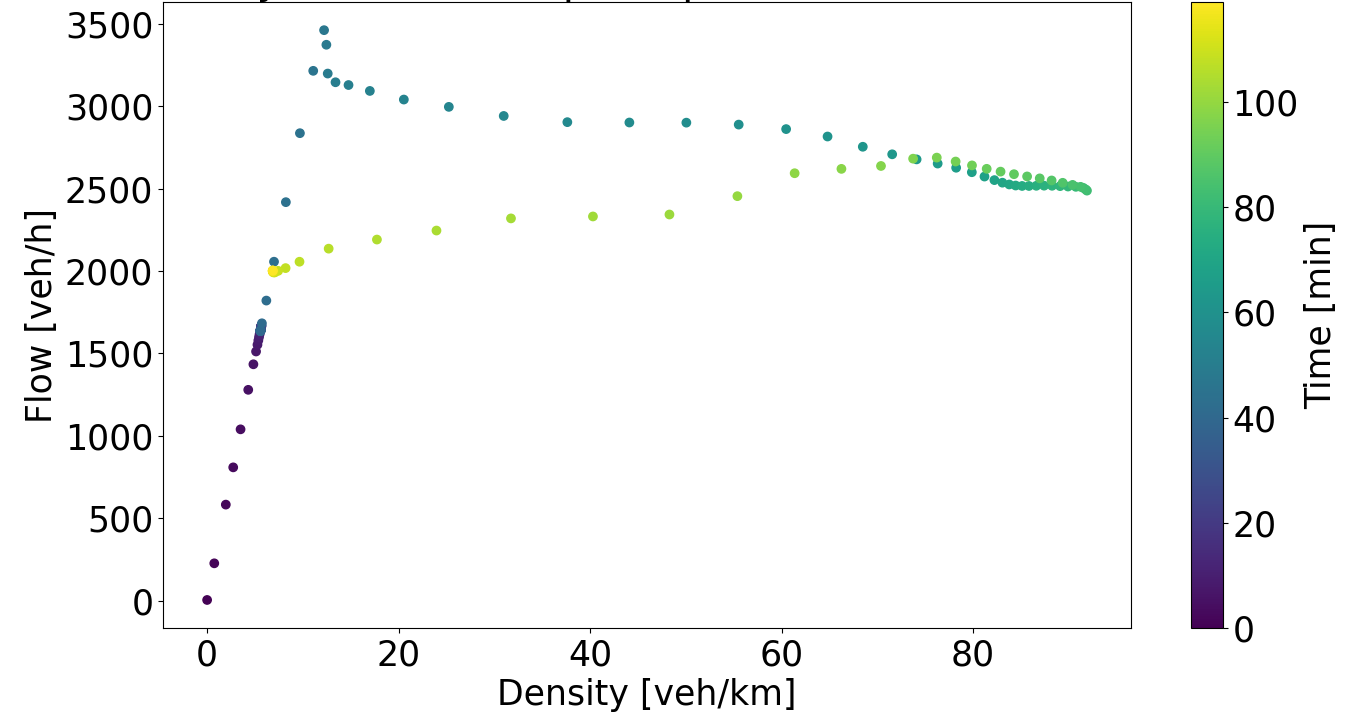}
        \caption{}
        \label{fig:link2lane}
    \end{subfigure}

    \caption{Hysteresis loop on the free-flow branch of MFD and its relation with the local bottleneck. (a) Hysteresis loop on the free-flow branch of MFD; (b) Density-outflow scatters when the local bottleneck is not triggered; (c) Density-flow states on the upstream of the bottleneck(3-lane segment) when there is local congestion; (d) Density-flow states on the upstream of the bottleneck(2-lane segment) when there is local congestion}
    \label{fig:hystall}
\end{figure}

Hysteresis loop in MFD is a phenomenon caused by different network performances before and after congestion. It usually happens near the capacity of the MFD \cite{Leclercq2019, Sirmatel2021, yuan-2023}, and is related to the density heterogeneity. The hysteresis loop is observed on the free-flow branch in the DF network (as shown before in Figure \ref{fig:mfddelft}). In the average flow MFD, a small hysteresis loop exists on the free-flow branch, as shown in Figure \ref{fig:hyst}. This is caused by the congestion at a lane drop on the freeway. For the three different demand levels, the flow-density evolution shows clear congestion on a three-lane link upstream of the bottleneck (Figure \ref{fig:link3lane}) and a two-lane link further upstream (Figure \ref{fig:link2lane}). The average outflow is influenced by the recovery of the local congestion, and hence the small hysteresis loop exists. 

This phenomenon has not been observed before to the best of the authors' knowledge. The MFD is usually used to describe an urban network with high connectivity. There are plenty of alternative routes and, hence, re-routing options. In this case, the congestion will naturally spread to other parts of the network, and the hysteresis loop will only appear when the network is saturated. However, in the DF network, the above-mentioned bottleneck is part of the only route for two OD pairs in high demand. To check if the bottleneck is indeed the cause of the hysteresis loop, a different OD matrix is tested where the demand of these two OD pairs is removed. As shown in Figure \ref{fig:remove bottleneck}, the hysteresis loop no longer exists in this situation. The hysteresis phenomenon and the inaccurate MFD estimation around the capacity in Figure \ref{fig:mfd} show the importance of a well-connected network to bathtub model studies. 

\section{Conclusions}
Different bathtub models can perform differently in different TDD and demand scenarios. Many model comparison studies have compared models in static TDD situations and assuming homogeneous trip distances or negative exponential TDD. In this study, we focused on the effect of TDD aggregation level and the time dependency of TDD on model performances. We compared the accumulation-based model, the M-model, and the trip-based models using event-based and fixed time step discretization with macroscopic traffic simulation data in two networks of Delft, one included the surrounding freeways, one focused on the urban roads, and a toy network. To do so, we first aggregated the macroscopic traffic simulation data to estimate the MFD and derive inputs for the bathtub models. We created 10 scenarios for the three networks with different demand profiles and TDD conditions. Through the comparison between scenarios, considerable influence of TDD aggregation level and time dependency was observed. We also found that the state transition speed of the networks can lead to different model performances. The accumulation-based model was the most accurate prediction in the two Delft networks, while trip-based models were more accurate in the toy network.

% The importance of the network structure and demand properties for MFD-based modeling were also shown in the simulation results. The low connectivity of both the Delft network and the toy network results in uneven distribution of congestion and makes MFD estimation inaccurate near the critical accumulation. The lack of alternative routes can also lead to the hysteresis loop on the free-flow branch of the average outflow MFD. For the small toy network and its evenly distributed demand over ODs, accumulation changes are more rapid and closer to the trip-based simulations. However, for the Delft network with unevenly distributed demand, the accumulation changes are slower and closer to the M model. 

The overall conclusions of this paper are:
\begin{itemize}
    \item We found that using less aggregated information of TDD in trip-based models can improve the accuracy of prediction, especially for the fast-varying demand case and TDD with high variation. Larger benefits in accuracy appear during the demand transition states. 
    \item The existence of dynamic TDD can increase the errors in the prediction of the bathtub models. The TDD transitions are more complex and harder to be accurately captured. In the two Delft networks, the accumulation-based model showed the best ability to adapt to dynamic TDDs.
    \item The state transition speed of the network can affect the prediction accuracy of different bathtub models. The slow response of the accumulation model can be compensated if the state transition is slow. Studies about the effect of network properties on bathtub models focused mainly on the MFD estimation \cite{geroliminis-2011, mousavizadeh-2024}. However, the difference in state transition speed can lead to different model performances when the MFD is fixed. This means that the state transition speed of networks should be considered in bathtub models instead of the MFD function.  
    \item In the two Delft networks, the accumulation-based model is more accurate in demand increasing stages than in decreasing stages. However, the trip-based models are more accurate in demand decreasing stages. This observation further supports the importance of studying network state transition speed and how to capture it in bathtub models. 
    % 
    % The network connectivity, road hierarchies, network size, and demand distribution over ODs can all affect the performance of MFD-based modeling. In case studies, these factors of the cases should be noted. 
    
\end{itemize}

We have also observed that the M-model performs worse than the accumulation-based model in every scenario, as the optimal alpha is always 0. 
% In the two Delft networks, this makes sense, as the slow network state transition speed makes the accumulation-based model the most accurate one. However, the M-model is still less accurate in the toy network scenarios than the accumulation-based model. 
This means that M-Model failed as an extension version of the accumulation-based model, and the ex post optimization of the parameter cannot provide a more accurate approximation. In future research about the M-model, it is necessary to understand this kind of failure and its relation with model parameters such as length of the time step and coefficient of variation of TDDs. 

The main limitation of this study is that a relativity low demand is assumed, and the saturated states are not discussed, which may be more valuable for further management and control design. Moreover, the data used in this study was not calibrated due to the lack of real-life data. Finally, the effect of re-routing in the change of TDD over time is not explicitly considered. It may also considerably influence TDD and, hence, the performance of bathtub models.

% We haven't discussed much about this. But it may be worth adding this to the conclusion. The M-model was developed as an approximation to aim to capture better the dynamics when the trip distance is not NE. However, the fact that the accumulation based model performs better than the M-model, suggest that such an approximation/enhancement to the model is not necessarily good.

For future research, the effect of dynamic TDD during saturation states is an important direction to explore. For saturated conditions, the re-routing of trips is expected to be more significant and influential on TDDs. Thus, it is essential to explicitly capture the re-routing effect to quantify the impacts. For networks with local bottlenecks, an extension in MFD estimation or dynamic partitioning strategy is needed to consider uneven congestion distributions. A further step is to explore to what extent the errors in bathtub models can affect the performance of control strategies. 

% For the application side, the most bathtub model should be the most suitable for specific network properties and control strategies. 

\bibliographystyle{trb}
\bibliography{trb_template}
\end{document}